\documentclass[aps,graphicx,reprint,amsmath,amssymb,nofootinbib,noeprint,longbibliography]{revtex4-1}

\usepackage[locale=US]{siunitx}
\usepackage{hyperref}
\usepackage{orcidlink}
\usepackage[capitalise]{cleveref}
\usepackage{bm}
\usepackage[]{tikz}
\usetikzlibrary{calc}
\usepackage{mathptmx}

\newcommand{\round}[2]{\num[round-mode=places,round-precision=#1]{#2}}

\DeclareSIUnit\angstrom{\text {Å}}

\definecolor{fzjblue}{HTML}{023d6b}
\definecolor{fzjlightblue}{HTML}{adbde3}
\definecolor{fzjgrey}{HTML}{ebebeb}
\definecolor{fzjgreen}{HTML}{b9d25f}
\definecolor{fzjyellow}{HTML}{faeb5a}
\definecolor{fzjviolet}{HTML}{af82b9}
\definecolor{fzjred}{HTML}{eb5f73}
\definecolor{fzjorange}{HTML}{fab45a}

\definecolor{bda_grey}{HTML}{B3B3B3}
\definecolor{bda_white}{HTML}{FFFFFF}
\definecolor{bda_lightblue}{HTML}{006DB3}
\definecolor{bda_darkblue}{HTML}{0414A8}
\definecolor{bda_orange}{HTML}{FFAA33}
\definecolor{bda_green}{HTML}{00FF00}
\definecolor{bda_purple}{HTML}{DF85FF}

\definecolor{dxa_bcc_green}{HTML}{00ff00}
\definecolor{dxa_bcc_pink}{HTML}{ff4dcc}
\definecolor{dxa_bcc_blue}{HTML}{3380ff}
\definecolor{dxa_bcc_red}{HTML}{e63333}

\begin{document}

\title{Nanoindentation simulations for copper and tungsten with adaptive-precision potentials}

\author{David Immel\,\orcidlink{0000-0001-5143-8043}}
\affiliation{Jülich Supercomputing Centre (JSC), Institute for Advanced Simulation (IAS), Forschungszentrum Jülich, Jülich, Germany}

\author{Matous Mrovec\,\orcidlink{0000-0001-8216-2254}}
\affiliation{Interdisciplinary Centre for Advanced Materials Simulations (ICAMS), Ruhr Universität Bochum, Bochum, Germany}

\author{Ralf Drautz\,\orcidlink{0000-0001-7101-8804}}
\affiliation{Interdisciplinary Centre for Advanced Materials Simulations (ICAMS), Ruhr Universität Bochum, Bochum, Germany}

\author{Godehard Sutmann\,\orcidlink{0000-0002-9004-604X}}
\email[]{g.sutmann@fz-juelich.de}
\affiliation{Jülich Supercomputing Centre (JSC), Institute for Advanced Simulation (IAS), Forschungszentrum Jülich, Jülich, Germany\\Interdisciplinary Centre for Advanced Materials Simulations (ICAMS), Ruhr Universität Bochum, Bochum, Germany}

\date{\today}

\begin{abstract}
We perform nanoindentation simulations for both the prototypical face-centered cubic metal copper and the body-centered cubic metal tungsten with a new adaptive-precision description of interaction potentials including different accuracy and computational costs: We combine both a computationally efficient embedded atom method (EAM) potential and a precise but computationally less efficient machine learning potential based on the atomic cluster expansion (ACE) into an adaptive-precision (AP) potential tailored for the nanoindentation. The numerically expensive ACE potential is employed selectively only in regions of the computational cell where large accuracy is required. The comparison with pure EAM and pure ACE simulations shows that for Cu, all potentials yield similar dislocation morphologies under the indenter with only small quantitative differences. In contrast, markedly different plasticity mechanisms are observed for W in simulations performed with the central-force EAM potential compared to results obtained using the ACE potential which is able to describe accurately the angular character of bonding in W due to its half-filled d-band. All ACE-specific mechanisms are reproduced in the AP nanoindentation simulations, however, with a significant speedup of 20-30 times compared to the pure ACE simulations. Hence, the AP potential overcomes the performance gap between the precise ACE and the fast EAM potential by combining the advantages of both potentials.
\end{abstract}


\maketitle

\section{Introduction}
\label{sec::introduction}
Large-scale molecular dynamics (MD) simulations of nanoindentations \cite{doi:10.1126/science.248.4954.454,PhysRevA.42.5844} can provide valuable insights into microscopic mechanical behavior of metallic materials and reveal underlying mechanisms that govern initial stages of plastic deformation. To obtain trustworthy results it is essential to describe the atomic interactions with accurate and transferable interatomic potentials~\cite{PhysRevA.42.5844}. For free-electron metals with close-packed face-centered cubic (FCC) crystal structures, many-body central force approaches, such as the embedded atom method~\cite{eam, DAW1993251} or the Finnis-Sinclair potentials~\cite{eam}, usually provide sufficient accuracy. However, many transition metals with partially filled d-bands cannot be described properly by such central-force models~\cite{pettifor1995bonding, mrovec2004bond, mrovec2007bond, GROGER20085401}. These metals, crystallizing in hexagonal close-packed (HCP) or body-centered cubic (BCC) structures, require explicit descriptions of their angular-dependent interactions. 

Two prototypical elements representing these families of metals are FCC copper and BCC tungsten. Understanding mechanical properties of both metals at the nanoscale is highly relevant due to ever increasing miniaturization. The refractory metal W is also considered as plasma facing material (PFM) for next generation fusion reactors~\cite{tungsten_for_nuclear_fusion,tungsten_iter_plasma_facing_components,tungsten_divertor_plasma}.  As PFMs need to sustain  extreme operating conditions, understanding changes of their mechanical properties due to formation of large number of crystal defects under irradiation is of utmost importance.
Even though nanoindentation simulations have already provided valuable mechanistic insights into microscale plasticity for both Cu~\cite{hansson2015,sahputra2021temperature,SHINDE2022125559,HUANG2021110237} and W~\cite{simulation_nanoindentation_tungsten_low_temperature_indenter_types,tungsten_nanoindentation_different_potentials_roomtemperature,tungsten_nanoindentation_eam_all_temperatures,ZHU2024110738,10.1063/5.0191162,cryst13030469}, further investigations and validations with improved interatomic potentials are still needed, especially for BCC W.

In recent years, machine-learning (ML) potentials proved to be able to deliver an accuracy close to that of first-principles electronic-structure methods at a fraction of their computational cost~\cite{smith2021automated, pace, HODAPP2024112715,mishin_mpl_review, mortazavi2024recent}. Even though there exist ML potentials for Cu~\cite{pace} and W \cite{PhysRevB.99.184305,Nikoulis_2021,PhysRevMaterials.6.083801,cho2023atomistic,PhysRevB.108.054312}, their application to large-scale nanoindentation simulations is still limited. One contributing reason is that most ML potentials remain several orders of magnitude slower than the simple empirical potentials~\cite{pace, doi:10.1021/acs.jctc.2c01149,Stark_2024}. 

A possible solution how to maintain desired accuracy while improving computational efficiency is by coupling different interaction models. Typically, a more accurate but computationally demanding model is used for a small portion of the simulated system where decisive processes take place (for instance, bond breaking at a crack tip, surface reactions, or defect nucleation sites). An efficient but less accurate model is then applied for remaining parts of the simulated system where only small variations are expected (for instance, maintaining a long range elastic field or temperature gradients). A well-known and widely applied example is coupling of quantum-mechanical (QM) approaches with molecular mechanics (MM) which originated almost 50 years ago \cite{qm_mm_coupling_hard_coded_zones1}. Current QM/MM implementations are able to adjust both regions during simulations based on various criteria \cite{PhysRevLett.96.095505, KERDCHAROEN1996313, energy_based_switching_qmmm}.  When QM calculations are not required, the coupling can be done also for interatomic potentials with different levels of accuracy and computational cost~\cite{adaptive_precision_potentials, birks2025efficientaccuratespatialmixing, ml_ff_coupling_force_based, doi:10.1021/acs.jctc.0c01112}.  

A viable way to overcome the performance gap between ML and simple empirical potentials is via the adaptive-precision (AP) method~\cite{adaptive_precision_potentials}. In the AP method, the potential energy is a weighted average of the combined potentials depending on a switching parameter.  The switching parameter is calculated by a customizable detection mechanism so that the assignment of atoms to a particular region can be done in an automatized way. Thus, the method works autonomously and self adaptively in space and time. An important feature of the AP method is that the time integration of the equations of motion maintains the conservation of energy and momentum.

In this work, we apply the AP method to combine the speed of EAM potentials and the accuracy of ACE potentials in large-scale nanoindentation simulations. Furthermore, we extend the original approach to account properly for homogeneous nucleation of dislocations that takes place below the indenter in regions of highest shear stress \cite{REMINGTON2014378}. These events marking the onset of plasticity need to be simulated with the accurate ACE potential.

\section{Methods}
\subsection{Nanoindentation}
\label{sec::nanoindentation_setup}

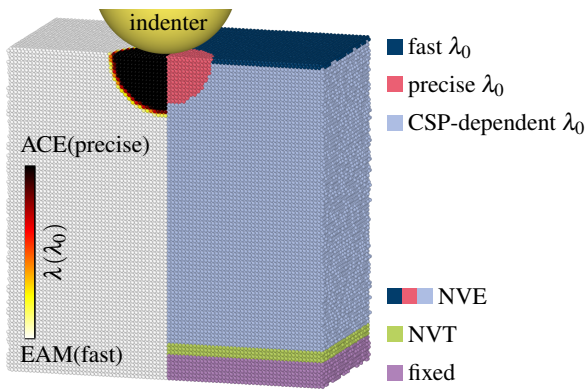
\begin{figure}[tb]
\newcommand{\myrule}[1]{\textcolor{#1}{\rule{0.2cm}{0.2cm}}}
\begin{tikzpicture}[x=2in,y=2in]
  \node[anchor=north west,inner sep=0] at (0,0) {
    \includegraphics[width=2in]{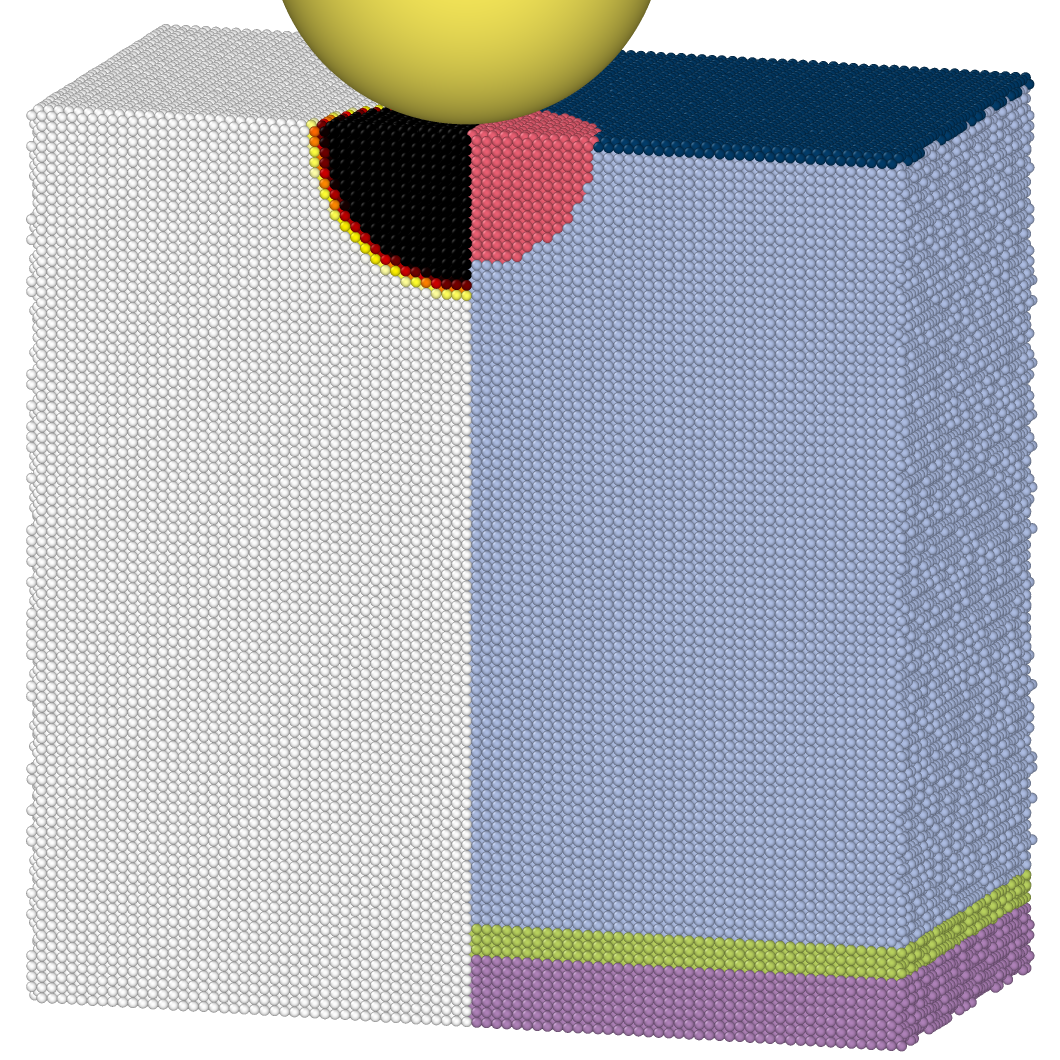}
  };
  \node[anchor=north] at (0.45, 0.01) {\textcolor{black}{indenter}};
  \node [coordinate] (cb_top) at (0.10,-0.41) [] {};
  \node [coordinate] (cb_bot) at (0.10,-0.86) [] {};
  \node[anchor=north east,inner sep=0] at (cb_top) {
    \frame{\includegraphics[height=0.9in]{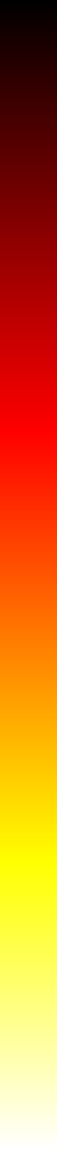}}
  };
  \node[anchor=south west] at ($(cb_top)-(0.06,0)$) {ACE(precise)};
  \node[anchor=north west] at ($(cb_bot)-(0.06,0)$) {EAM(fast)};
  \draw[] (cb_top) -- (cb_bot) node[midway, below, rotate=90] {$\lambda(\lambda_0)$};

  \node[anchor=west] at (1, -0.1) {\myrule{fzjblue} fast $\lambda_0$};
  \node[anchor=west] at (1, -0.2) {\myrule{fzjred} precise $\lambda_0$};
  \node[anchor=west] at (1, -0.3) {\myrule{fzjlightblue} CSP-dependent $\lambda_0$};

  \node[anchor=west] at (1, -0.75) {\myrule{fzjblue}\myrule{fzjred}\myrule{fzjlightblue} NVE};
  \node[anchor=west] at (1, -0.85) {\myrule{fzjgreen} NVT};
  \node[anchor=west] at (1, -0.95) {\myrule{fzjviolet} fixed};
\end{tikzpicture}
\caption{\label{Fig::vis_nanoindentation_setup}
Cut through the system along the (100) plane below the indenter.
Right side: The top atoms (\myrule{fzjblue}\myrule{fzjred}\myrule{fzjlightblue}) are simulated in a NVE ensemble, the bottom atoms are fixed (\myrule{fzjviolet}), and the atoms in between are simulated in a NVT ensemble (\myrule{fzjgreen}).
For simulations with an adaptive-precision potential, the parameter $\lambda_0$ is used to calculate the switching parameter $\lambda$ which is required to calculate the potential energy according to \cref{eq::energy_hyb}.
To ensure a precise simulation, $\lambda_0=0$ is set after the equilibration for all atoms which are in a hemisphere (\myrule{fzjred}) whose centre is on the surface of the material under the centre of the indenter.
The centro-symmetry parameter (CSP) is used to calculate $\lambda_0$ for most atoms (\myrule{fzjlightblue}) to dynamically adjust the precision.
To save the computation time, $\lambda_0=1$ is set at the surface outside of the hemisphere (\myrule{fzjblue}) to prevent the detection of the surface by the CSP.
Left side: Initially used switching parameter $\lambda$ calculated from $\lambda_0$.
A region of interest below the indenter is calculated precisely while the remaining atoms are calculated quickly.
}
\end{figure}

The MD simulations were performed using the LAMMPS software package~\cite{lammps}. One of common approaches in the nanoindentation simulations is to separate the simulation block into three parts~\cite{tungsten_nanoindentation_eam_all_temperatures, tungsten_nanoindentation_different_potentials_roomtemperature, relaxation_nanoindentation_mysource}, as visualized in \cref{Fig::vis_nanoindentation_setup}. The bottom part is kept fixed to prevent rotation and translation of the sample, the central part is simulated using the NVT ensemble to dissipate heat generated by the indentation, and the top part is simulated using the NVE ensemble.  The spherical rigid indenter was described by the following potential~\cite{csp}
\begin{equation}
V_\text{ind}(\vec{r}) = \frac{k_\text{ind}}{3} \Theta(R_\text{ind}-\lvert\vec{r}_\text{ind}-\vec{r}\rvert)(R_\text{ind}-\lvert\vec{r}_\text{ind}-\vec{r}\rvert)^3\,,
\label{eq::V_indenter}
\end{equation}
where $k_\text{ind}$ is the force constant, $\vec{r}_\text{ind}(t)$ the center of the indenter, $R_\text{ind}$ the indenter radius, and $\Theta(x)$ the Heaviside step function. In the beginning of the simulation, the indenter and surface are not in contact and the center of the indenter is located at $\vec{r}_{\text{ind},0}$. The indenter is then gradually inserted into the sample with a constant velocity $\vec{v}_\text{ind}$ such that its position changes as $\vec{r}_\text{ind}(t) = \vec{r}_{\text{ind},0} - \vec{v}_\text{ind} t$.
Further details about simulation setup are provided in \cref{sec::equilibration}.

\subsection{Interatomic potentials}
\label{sec::interatomic_potentials}

\subsubsection{EAM potential}
In the EAM potentials used in this work,  the energy of an atom $i$ is given as
\begin{equation}
E_i^\text{EAM} = \xi\left(\sum_{j\neq i}\zeta(r_{ij})\right) + \frac{1}{2} \sum_{j\neq i} \Phi(r_{ij}),
\label{eq::energy_eam}
\end{equation}
where $\xi$ is the embedding function, $\zeta$ is the electron charge density and $\Phi$ is the pair potential.
For Cu, we used the EAM2 parametrizations of Ref. \cite{mishin} which has been thoroughly tested and widely used~\cite{mishin_use_1,mishin_use_2,mishin_use_3,mishin_use_4}. For W, we used the EAM2 potential from Ref. \cite{eam_tungsten_2013} which was developed for simulating radiation defects and dislocations~\cite{Bonny_2014,ZHOU2014202,tungsten_nanoindentation_different_potentials_roomtemperature,stupak2020structure}.

\subsubsection{Atomic cluster expansion}
The employed ACE potentials used a combination of linear and square-root expansions that were shown to give the best accuracy and computational performance~\cite{pace}. 
The potential energy of an atom $i$ described by ACE is given as
\begin{equation}
E_i^\text{ACE} = \phi_i^{(1)} + \sqrt{\phi_i^{(2)}}\,.
\end{equation}
The functions $\phi_i^{(p)}$ are expanded as
\begin{equation}
\phi_i^{(p)} = \sum_{\bm{v}} c_{\bm{v}}^{(p)} B_{i\bm{v}}\,,
\label{eq::energy_ace}
\end{equation}
where $B_{i\bm{v}}$ are product basis functions, which describe the atomic environment, and $c_{\bm{v}}^{(p)}$ are fitted expansion coefficients with the multi-indices $\bm{v}$. For Cu, we used the ACE parametrization developed in Ref.~\cite{pace} which has been applied to dislocation simulations~\cite{NAMAKIAN2023111971}. For W, we used an ACE parametrization trained on an extensive DFT dataset of bulk structures as well as defects. Further information of this potential is provided in \cref{sec::ace_tungsten_potential}.

\subsubsection{Adaptive-precision potential}
\label{sec::adaptive_precision_potential}
The atomic energy of the AP potential~\cite{adaptive_precision_potentials}, which combines EAM and ACE, is given by
\begin{equation}
E_i = \lambda_i E_i^\text{EAM} + (1-\lambda_i) E_i^\text{ACE}\,,
\label{eq::energy_hyb}
\end{equation}
where  $\lambda_i\in[0,1]$ is the switching parameter~\cite{adaptive_precision_potentials}. We used the switching parameter based on the centro-symmetry parameter (CSP) \cite{csp} to detect atoms which need to be treated with the more accurate potential. In addition, we fixed the value of $\lambda_i$ in specific regions of the simulation block, as visualized in \cref{Fig::vis_nanoindentation_setup}.  Since the first dislocations nucleate primarily in the regions of highest shear stress near the indenter surface~\cite{REMINGTON2014378}, the region below the indenter is always simulated with the ACE potential. In practice, we set the initial value of $\lambda_{0,i}=0$ in a hemisphere of radius  $\SI{40}{\angstrom}$ whose center is at the contact point where the indenter first touches the surface.  Additionally, to prevent the CSP from detecting surface atoms, we set $\lambda_{0,i}=1$ for surface atoms which are outside of the ACE hemisphere. Using this setup, we minimize the amount of the computationally more expensive ACE calculations. During later stages of the simulation, $\lambda_{i}$ is recalculated and updated for the remaining atoms in the block depending on the centro-symmetry parameter~\cite{adaptive_precision_potentials}. Details of the AP potentials are provided in \cref{sec::hybrid_tungsten_potential}.

\section{Results}
\subsection{Copper}
\label{sec::results_copper}
\begin{table}
\caption{\label{tab::setup_parameters_copper}
Parameter values used in the copper nanoindentation simulations.
}
\centering
\begin{tabular}{lr}
\hline\hline
parameter & value\\\hline
surface orientation & $(100)$\\
temperature & $\SI{292}{\kelvin}$\\
simulation box size & $\SI{364.2}{\angstrom} \times \SI{364.2}{\angstrom} \times \SI{363.2}{\angstrom}$\\
number of atoms & $4, 000, 000$\\
height of fixed region & $\SI{10}{\angstrom}$\\
height of NVT region & $\SI{10}{\angstrom}$\\
timestep $\Delta t$ & $\SI{1}{\femto\second}$\\
indenter radius $R_\text{ind}$ & $\SI{60}{\angstrom}$\\
indenter velocity $\vec{v}_\text{ind}$ & $\{0,0,25\}\,\si{\metre/\second}$\\
indenter force constant $k_\text{ind}$ & $\SI{10}{\electronvolt\angstrom^{-3}}$\\
indentation depth $h_\text{max}$ & $\SI{20}{\angstrom}$\\
initial indentation depth $h_0$ & $\SI{-5}{\angstrom}$\\
\hline\hline
\end{tabular}
\end{table}

Before staring the nanoindentation simulation, we equilibrate the Cu$_\text{(100)}$ surface as described in \cref{sec::equilibration}. The nanoindentation simulation is then performed using the $NVE$ ensemble, but a thin layer of atoms at the bottom of the simulation box is thermostated and the very bottom layers are kept fixed to dissipate heat generated during the indentation and to prevent translation and rotation of the sample, as visualized in \cref{Fig::vis_nanoindentation_setup}.
The fundamental parameters of the simulation are summarized in \cref{tab::setup_parameters_copper}. 

\begin{figure}[tb]
\includegraphics[width=3.37in]{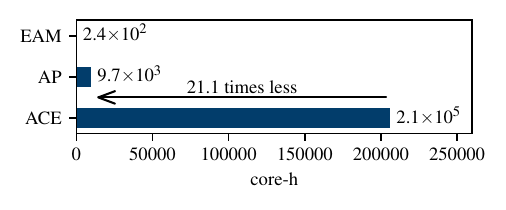}
\caption{\label{Fig::speedup_nanoindentation_copper}
Total computation time of a nanoindentation with 4 million Cu atoms simulated for $\SI{100}{\pico\second}$ with the AP potential compared to ACE and EAM simulations. EAM, AP and ACE simulations are calculated on 128, 128 and 2048 cores of JURECA-DC\cite{jureca}, respectively.
}
\end{figure}

The nanoindentation simulations were carried out on the JURECA-DC supercomputer~\cite{jureca}.
The speedup of the AP nanoindentation simulation compared to that performed with ACE only is about 21, as shown in \cref{Fig::speedup_nanoindentation_copper}.
The fixing of the switching parameter in the zones of interest, as described in \cref{sec::adaptive_precision_potential}, results in a speedup of 1.9 of the AP simulation compared to the AP simulation in Ref. \cite{adaptive_precision_potentials}.
Further details of the computational efficiency are provided in \cref{sec::computational:efficiency}.

\begin{figure}[tb]
\begin{center}

\begin{tikzpicture}[x=1.1in,y=1.1in]
  \node[anchor=north west,inner sep=0] at (0,0) {
    \includegraphics[width=1.1in]{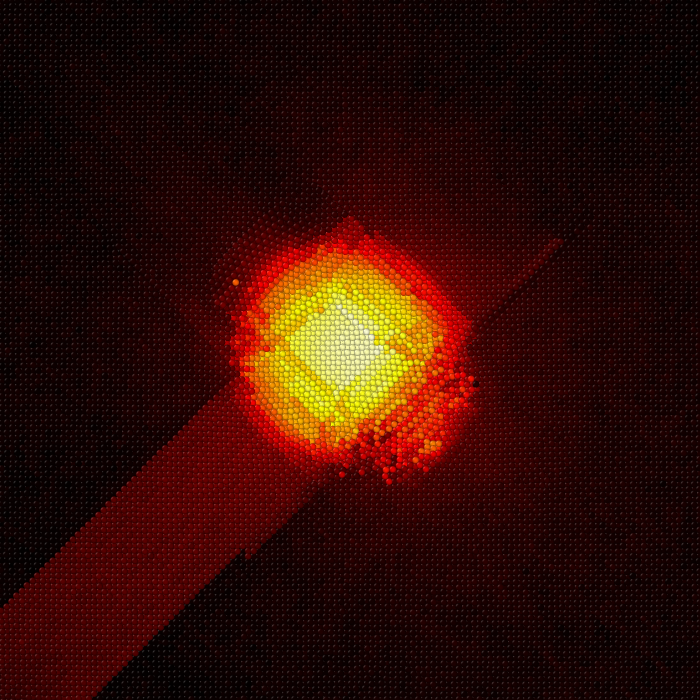}
  };
  \node[anchor=north] at (0.5,0.0) {\textcolor{white}{EAM}};
  \node[anchor=north west] at (0,0) {\textcolor{white}{a)}};
  \node [coordinate] (tripod_origin) at (0.15,-0.90) [] {};
  \draw[white,->] (tripod_origin) -- ($(tripod_origin) + (0.1,0)$) node[anchor=west]  {$[100]$};
  \draw[white,->] (tripod_origin) -- ($(tripod_origin) + (0,0.1)$) node[anchor=south] {$[010]$};
\end{tikzpicture}
\begin{tikzpicture}[x=1.1in,y=1.1in]
  \node[anchor=north west,inner sep=0] at (0,0) {
    \includegraphics[width=1.1in]{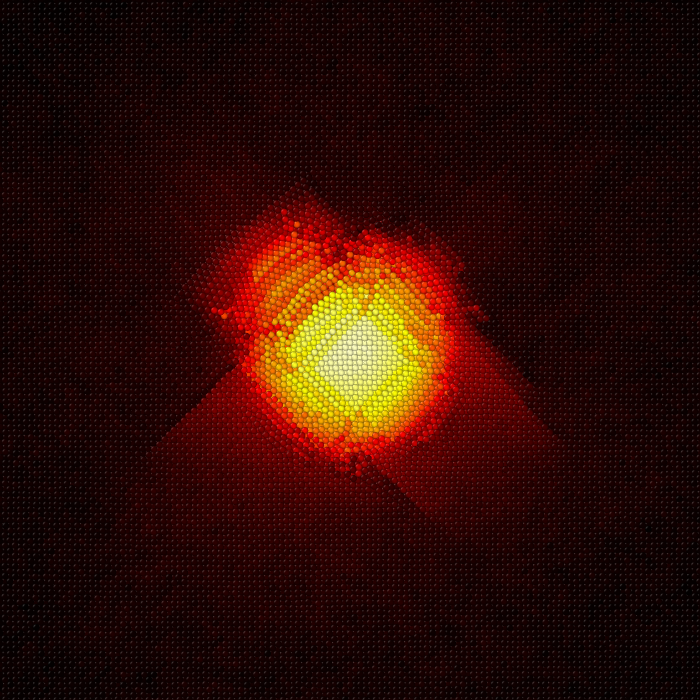}
  };
  \node[anchor=north] at (0.5,0.0) {\textcolor{white}{AP}};
  \node[anchor=north west] at (0,0) {\textcolor{white}{b)}};
\end{tikzpicture}
\begin{tikzpicture}[x=1.1in,y=1.1in]
  \node[anchor=north west,inner sep=0] at (0,0) {
    \includegraphics[width=1.1in]{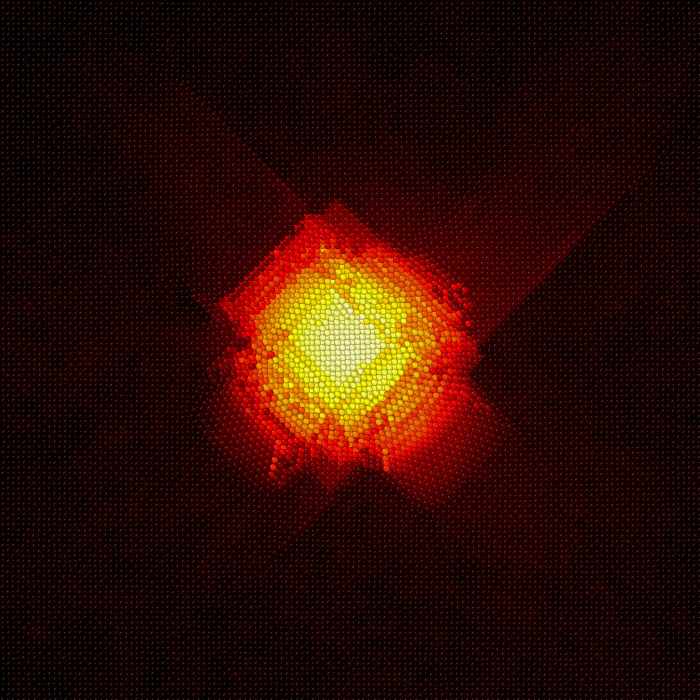}
  };
  \node[anchor=north] at (0.5,0.0) {\textcolor{white}{ACE}};
  \node[anchor=north west] at (0,0) {\textcolor{white}{c)}};
\end{tikzpicture}
\begin{tikzpicture}[x=1.000in,y=0.050in]
  \node[anchor=north west,inner sep=0] at (0,0) {
    \frame{\includegraphics[width=1in]{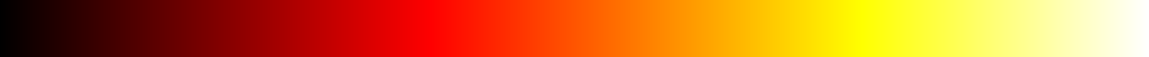}}
  };
  \node[anchor=south] at (0.5,0.0) {$\Delta r / \si{\angstrom}$};
  \node[anchor=east] at (0.0,-0.5) {0};
  \node[anchor=west] at (1.0,-0.5) {22};
\end{tikzpicture}
 \includegraphics[width=\columnwidth]{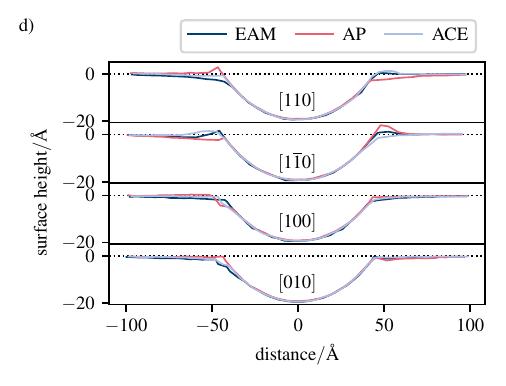}
\end{center}
\caption{\label{Fig::displacement_surface_copper}
a-c) Displacement $\Delta r$ of the copper atoms at $\SI{20}{\angstrom}$ indentation depth compared with $\SI{-5}{\angstrom}$ indentation depth.
d) Cross sections through the center of the supercell below the indenter along different directions at $\SI{20}{\angstrom}$ indentation depth.
}
\end{figure}
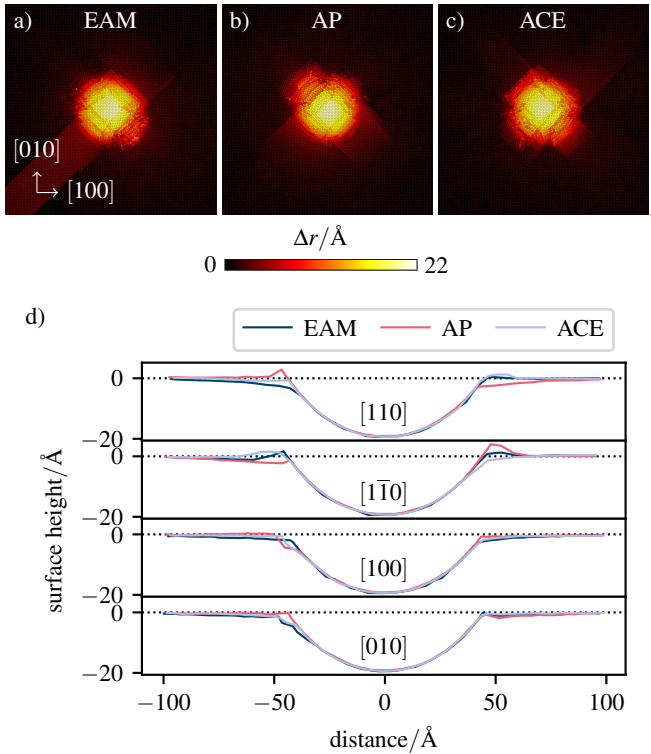

The displacement of the surface atoms which developed during the nanoindentations with the three potentials is visualized in \cref{Fig::displacement_surface_copper}(a-c).
The height of the surface on a line through the contact point of surface and indenter is shown for cross sections along the $\langle 110 \rangle$ and $\langle 100 \rangle$ directions in \cref{Fig::displacement_surface_copper}(d).
We observe small pile-ups in the $\langle 110 \rangle$ directions, similarly as in Ref. \cite{SHINDE2022125559}, and no pile-up in the $\langle 100 \rangle$ directions.

\begin{figure}[tb]
\begin{center}
 \includegraphics[width=3.3in]{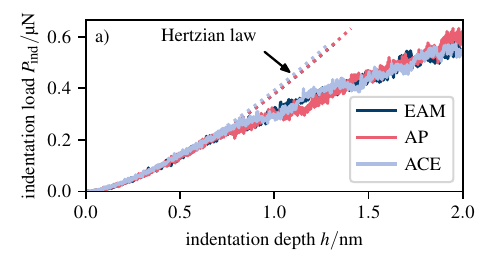}
 \includegraphics[width=3.3in]{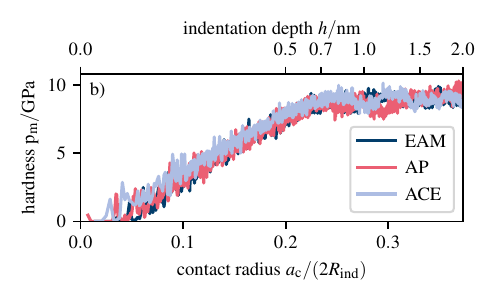}
\end{center}
\caption{\label{Fig::hertz_fit_copper}
a) Indentation load $P_\text{ind}$ and elastic Hertzian load $P_\text{H}$ according to \cref{eq::P_ind,eq::hertz_fit} computed for the copper nanoindentations.
b) Mean contact pressure or hardness $p_\text{m}$ according to \cref{eq::p_m} dependent on the contact radius $a_\text{c}$ according to \cref{eq::contact_radius}.
The contact radius is normalized with the radius $R_\text{ind}$ of the spherical indenter.
}
\end{figure}

The force between an atom $i$ and the indenter is given as $\vec{F}_\text{ind}(\vec{r}_i)=-\vec{\nabla}_{|\vec{r}_\text{ind}-\vec{r}|} V_\text{ind}(\vec{r}_i)$ with $V_\text{ind}$ according to \cref{eq::V_indenter}.
The indentation load
\begin{equation}
\vec{P}_\text{ind} = \sum_i \vec{F}_\text{ind}(\vec{r}_i)\cdot \frac{\vec{v}_\text{ind}}{v_\text{ind}}
\label{eq::P_ind}
\end{equation}
is given as total force on all atoms in the indentation direction $[00\overline{1}]$ and automatically calculated during the simulation.
The Hertzian analysis \cite{Hertz_elastische_kugeln, nanoindentation_analysis_hertz_1, nanoindentation_analysis_hertz_2} is applied to evaluate the Hertzian load $P_\text{H}$ on the surface as a function of the indentation depth $h$ as
\begin{equation}
P_\text{H} = \frac{4}{3} E^{*} R_\text{ind}^{1/2} h^{3/2},
\label{eq::hertz_fit}
\end{equation}
where $R_\text{ind}$ is the indenter radius and $E^{*}$ the indentation modulus.
This formula does not consider elastic anisotropy of materials, but it holds for spherical indenters with a modified indentation modulus \cite{hertz_analysis_anisotropy_improvements1,hertz_analysis_anisotropy_improvements2,nanoindentation_analysis_hertz_2}.
The Hertz fit allows to detect the onset of plastic deformation.
Both the simulated load $P_\text{ind}$ and the fitted analytical solution $P_\text{H}$ are shown in \cref{Fig::hertz_fit_copper}(a).
The fits give $E^{*}_\text{EAM}=\SI{115.9}{\giga\pascal}$, $E^{*}_\text{AP}=\SI{116.1}{\giga\pascal}$ and $E^{*}_\text{ACE}=\SI{112.0}{\giga\pascal}$ and are thus similar for all potentials.

The hardness or mean contact pressure $p_\text{m}$ on the surface is
\begin{equation}
p_\text{m} = \frac{P_\text{ind}}{A_\text{c}} = \frac{P_\text{ind}}{\pi a_\text{c}^2}\,,
\label{eq::p_m}
\end{equation}
where $A_\text{c}$ is the projected contact area with the contact radius $a_\text{c}$.
There are different models for the contact radius \cite{SHINDE2022125559,sahputra2021temperature,dissertation_nanocontact_radius}.
We use the contact radius
\begin{equation}
a_\text{c} = \sqrt{h (2R_\text{ind}- h)}
\label{eq::contact_radius}
\end{equation}
as this model was applied in nanoindentation simulations of Cu \cite{SHINDE2022125559} and W \cite{tungsten_nanoindentation_eam_all_temperatures} with spherical rigid indenters and thus enables a seamless comparison of our results with the literature.
The mean contact pressure $p_\text{m}$ is shown in \cref{Fig::hertz_fit_copper}(b).
It increases linearly until plastic deformation sets in at $a/(2R_\text{ind})\approx 0.23$ and reaches a steady state for larger contact radii.
This behavior is independent of the interatomic potential used, as expected from Refs. \cite{tungsten_nanoindentation_different_potentials_roomtemperature,dissertation_nanocontact_radius}.
The average hardness at the steady state is $\SI{8.9}{\giga\pascal}$ with a standard deviation up to $\SI{0.5}{\giga\pascal}$.
The hardness reported for $\text{Cu}_{(100)}$ in Ref. \cite{SHINDE2022125559} is $\SI{11.7\pm1.4}{\giga\pascal}$ and thus somewhat higher, but the indenter velocity was two times faster than in our case; as the hardness increases with the indenter velocity \cite{Imran_2012,Fang_2002}, a larger hardness magnitude is therefore expected.

\begin{figure}[tb]
\begin{tikzpicture}[x=1.000in,y=0.730in]
  \node[anchor=north west,inner sep=0] at (0,0) {
    \frame{\includegraphics[width=1in]{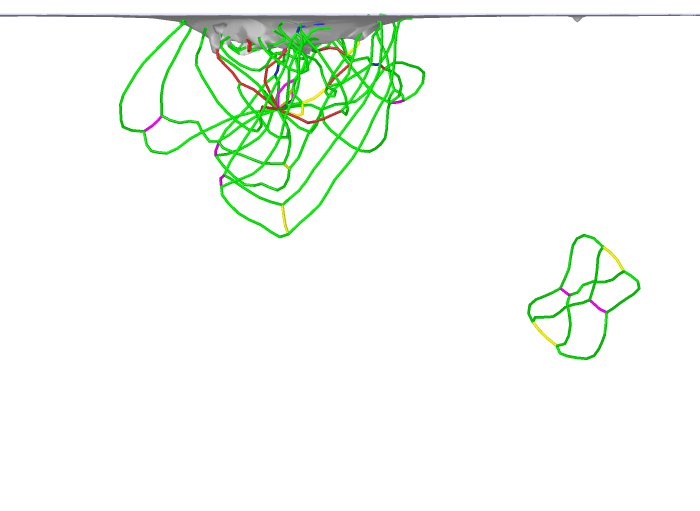}}
  };
  \node[anchor=south] at (0.5,0.0) {EAM};
  \node[anchor=north west] at (0,-0.03) {\textcolor{black}{a)}};
  \node [coordinate] (tripod_origin) at (0.60,-0.85) [] {};
  \draw[->] (tripod_origin) -- ($(tripod_origin) - (0.1,0)$) node[anchor=east]  {$[100]$};
  \draw[->] (tripod_origin) -- ($(tripod_origin) + (0,0.166)$) node[anchor=south] {$[001]$};
\end{tikzpicture}
\begin{tikzpicture}[x=1.000in,y=0.730in]
  \node[anchor=north west,inner sep=0] at (0,0) {
    \frame{\includegraphics[width=1in]{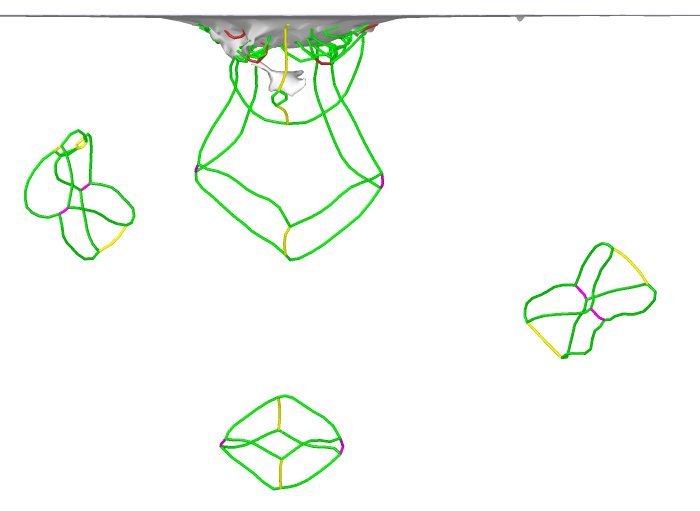}}
  };
  \node[anchor=south] at (0.5,0.0) {AP};
  \node[anchor=north west] at (0,-0.03) {\textcolor{black}{b)}};
\end{tikzpicture}
\begin{tikzpicture}[x=1.000in,y=0.730in]
  \node[anchor=north west,inner sep=0] at (0,0) {
    \frame{\includegraphics[width=1in]{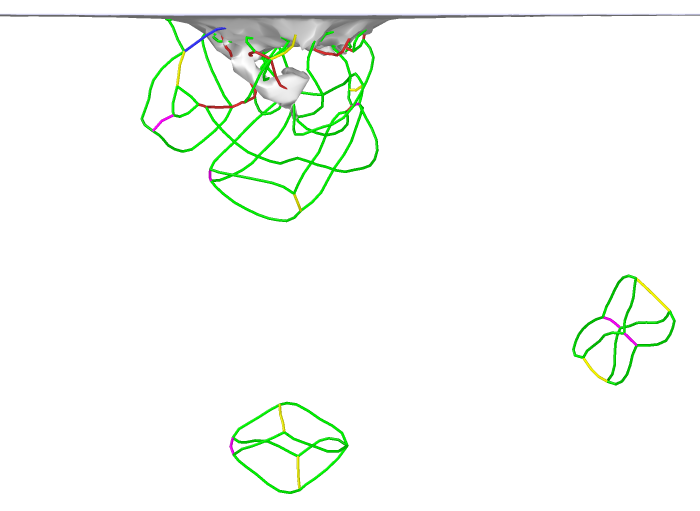}}
  };
  \node[anchor=south] at (0.5,0.0) {ACE};
  \node[anchor=north west] at (0,-0.03) {\textcolor{black}{c)}};
  \draw[|-|] (1.05,0) -- (1.05,-1) node[midway, above, rotate=270] {$\SI{181.5}{\angstrom}$};
\end{tikzpicture}
\begin{tikzpicture}[x=1.000in,y=1.473in]
  \node[anchor=north west,inner sep=0] at (0,0) {
    \frame{\includegraphics[width=1in]{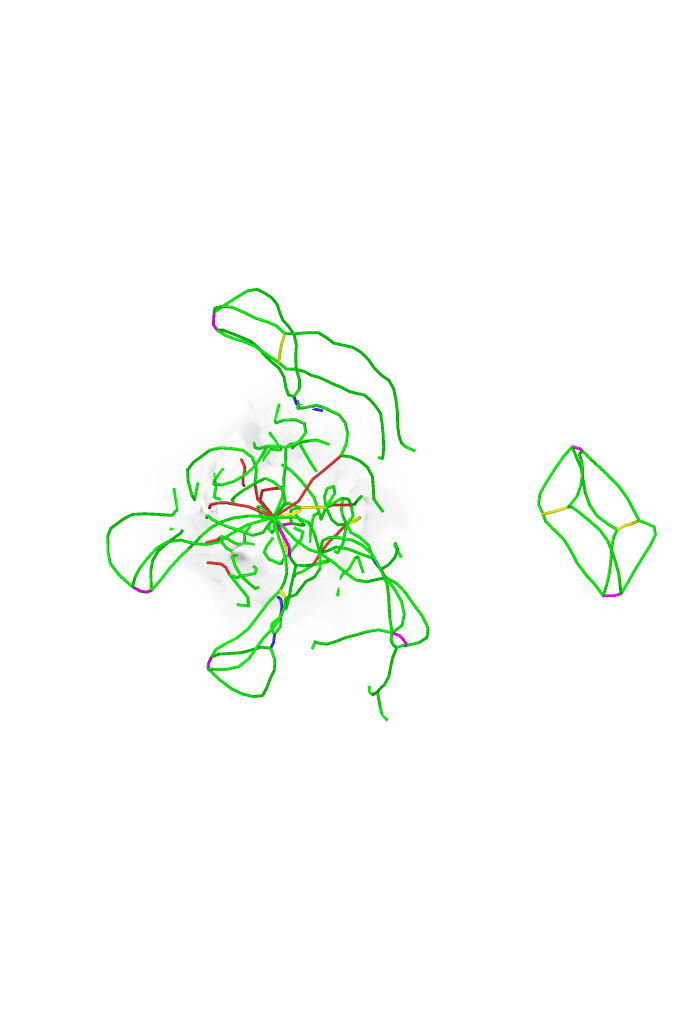}}
  };
  \node[anchor=north west] at (0,-0.3) {\textcolor{black}{d)}};
  \node [coordinate] (tripod_origin) at (0.80,-0.92) [] {};
  \draw[->] (tripod_origin) -- ($(tripod_origin) - (0.1,0)$) node[anchor=east]  {$[100]$};
  \draw[->] (tripod_origin) -- ($(tripod_origin) + (0,0.105)$) node[anchor=south] {$[010]$};
\end{tikzpicture}
\begin{tikzpicture}[x=1.000in,y=1.473in]
  \node[anchor=north west,inner sep=0] at (0,0) {
    \frame{\includegraphics[width=1in]{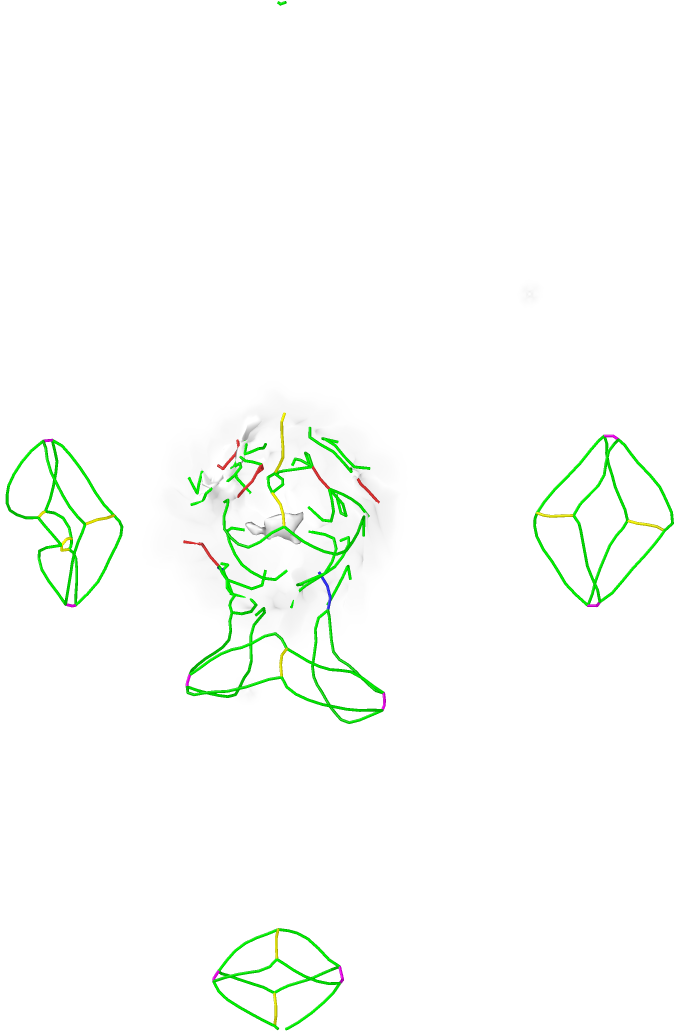}}
  };
  \node[anchor=north west] at (0,-0.3) {\textcolor{black}{e)}};
\end{tikzpicture}
\begin{tikzpicture}[x=1.000in,y=1.473in]
  \node[anchor=north west,inner sep=0] at (0,0) {
    \frame{\includegraphics[width=1in]{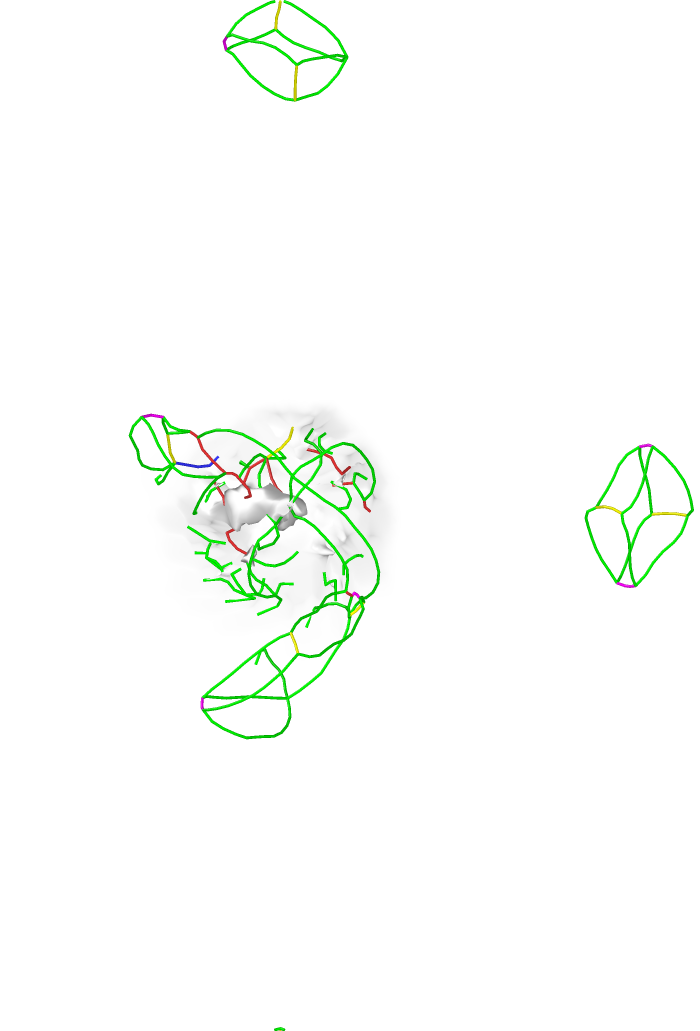}}
  };
  \node[anchor=north west] at (0,-0.3) {\textcolor{black}{f)}};
  \draw[|-|] (1.05,0) -- (1.05,-1) node[midway, above, rotate=270] {$\SI{365.9}{\angstrom}$};
\end{tikzpicture}
\begin{tikzpicture}[x=3.300in,y=0.111in]
  \node[anchor=north west,inner sep=0] at (0,0) {
    \includegraphics[width=3.3in]{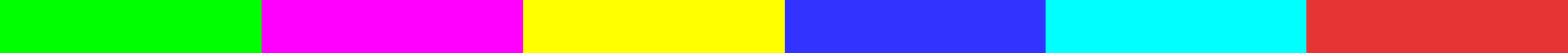}
  };
  \node[anchor=north] at (0.083,-1.0) {$1/6\langle112\rangle$};
  \node[anchor=north] at (0.250,-1.0) {$1/6\langle110\rangle$};
  \node[anchor=north] at (0.417,-1.0) {$1/3\langle100\rangle$};
  \node[anchor=north] at (0.583,-1.0) {$1/2\langle110\rangle$};
  \node[anchor=north] at (0.750,-1.0) {$1/3\langle111\rangle$};
  \node[anchor=north] at (0.916,-1.0) {other};
  \node[anchor=south] at (0.5,0.0) {dislocation type};
\end{tikzpicture}
\caption{\label{Fig::dislocation_networks_copper}
Dislocation lines in Cu identified by the DXA analysis of OVITO for a normalized indentation depth of $h/a_\text{c}\approx0.34$.
The dislocation lines are color-coded according to their identified dislocation type.
A scale bar is given on the right side for every row.
}
\end{figure}
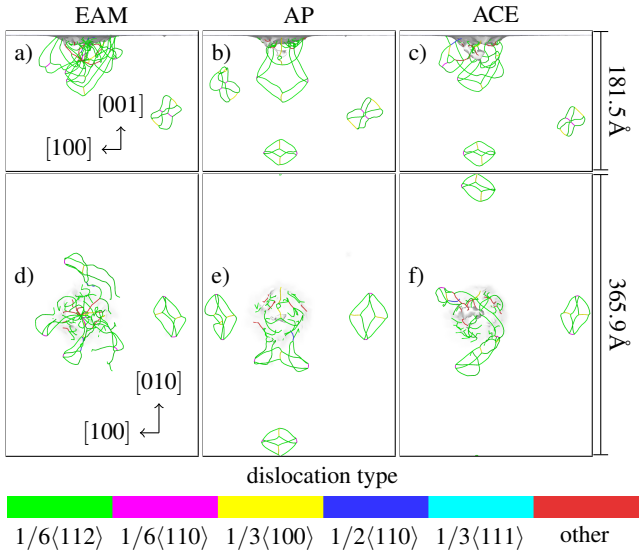

The dislocation extraction algorithm (DXA) \cite{ovito_dxa} implemented in OVITO \cite{ovito} was used to calculate the dislocation length $l_{klm}$ of dislocations of type $klm$.
For the FCC lattice, DXA identifies dislocations with the Burgers vectors $1/6\langle 112\rangle$, $1/6\langle 110\rangle$, $1/3\langle100\rangle$, $1/2\langle110\rangle$, $1/3\langle111\rangle$ and other Burgers vectors.
Dislocation networks observed in the simulation for the indentation depth $h/a_\text{c}\approx0.34$ are shown in \cref{Fig::dislocation_networks_copper}.

Prismatic dislocation loops nucleate like in Ref. \cite{SHINDE2022125559} for all potentials.
The prismatic dislocation loops glide in the directions $[10\overline{1}]$, $[\overline{1}0\overline{1}]$, $[01\overline{1}]$ and $[0\overline{1}\overline{1}]$.
Five dislocation loops nucleate during the simulation for EAM while four nucleate for the AP potential and ACE.

\begin{figure}[tb]
\begin{center}
 \includegraphics[width=\columnwidth]{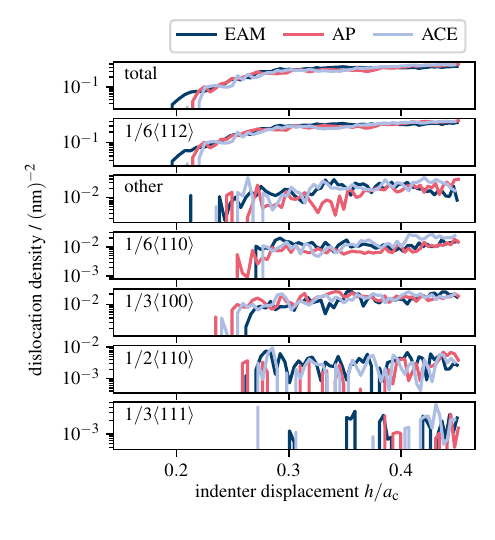}
\end{center}
\caption{\label{Fig::dislocation_density_copper}
Dislocation density according to \cref{eq::dislocation_density} for nanoindentations of Cu$_\text{(100)}$.
The dislocation density is given for all by the DXA of OVITO identified dislocation types.
Furthermore, the total type-independent dislocation density is shown.
}
\end{figure}

The dislocation density
\begin{equation}
\label{eq::dislocation_density}
\rho_{klm}^\text{dxa} = l_{klm} / V^\text{dxa}
\end{equation}
is calculated within the effective volume $V^\text{dxa}$ of the plastic zone.
The radius $a_\text{pz}$ of the plastic zone is given by
\begin{equation}
a_\text{pz} = f_\text{pz} a_\text{c}\,,
\label{eq::a_pz}
\end{equation}
where the factor $f_\text{pz}\in[0,3.5]$ depends on the material \cite{durst_plastic_zone}.
For Cu, we used $f_\text{pz}=1.9$ 
\cite{wang_scaling_factor_plastic_zone_copper}.
Thus, the effective volume of the plastic zone is given by
$V^\text{dxa} = 2 \pi a^3_\text{pz}/3 - V_\text{ind}$,
where the volume $V_\text{ind}$ displaced by the indenter, without considering sink-in and pile-up effects, is given as
$V_\text{ind} = \pi h^2 (R_\text{ind}-h)/3$
\cite{tungsten_nanoindentation_different_potentials_roomtemperature,begau_v_ind}.
The densities of different dislocation types as a function of the indentation depth are shown in \cref{Fig::dislocation_density_copper}.
As expected, the first nucleated dislocation has the Burgers vector $1/6\langle112\rangle$ for all potentials, but the nucleation occurs earlier for EAM than for the AP potential and ACE. We observe all dislocation types being nucleated for all potentials.

\begin{figure}[tb]
\begin{tikzpicture}[x=1.000in,y=0.737in]
  \node[anchor=north west,inner sep=0] at (0,0) {
    \includegraphics[width=1in]{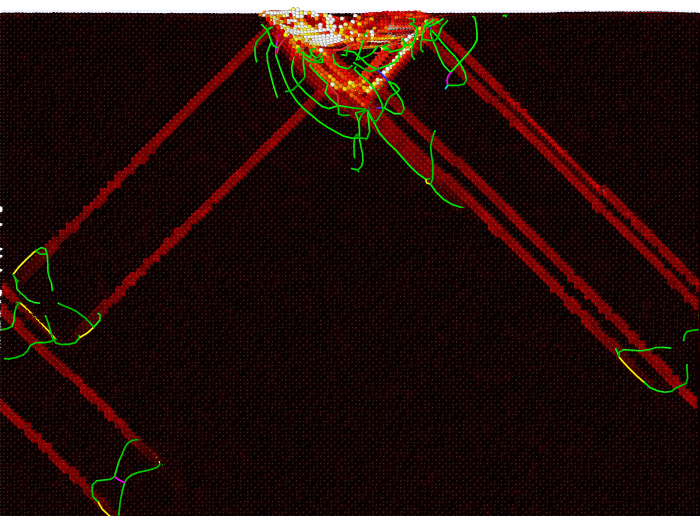}
  };
  \node[anchor=south] at (0.5,0.0) {EAM};
  \node[anchor=north west] at (0,0) {\textcolor{white}{a)}};
\end{tikzpicture}
\begin{tikzpicture}[x=1.000in,y=0.737in]
  \node[anchor=north west,inner sep=0] at (0,0) {
    \includegraphics[width=1in]{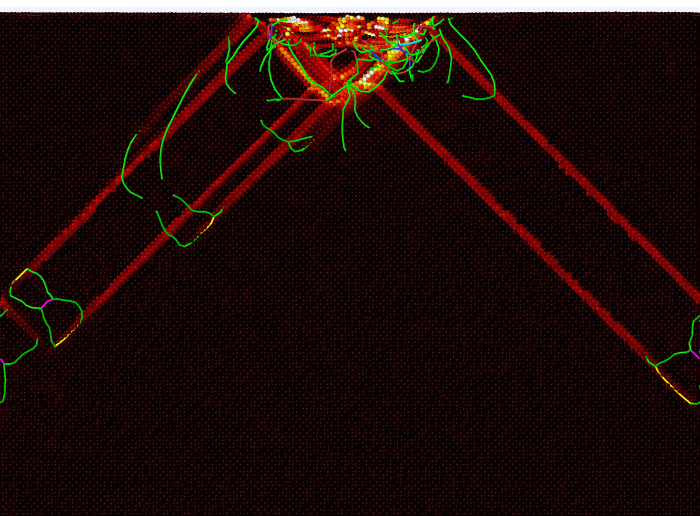}
  };
  \node[anchor=south] at (0.5,0.0) {AP};
  \node[anchor=north west] at (0,0) {\textcolor{white}{b)}};
  \node[anchor=south] at (0.6,-0.85) {\textcolor{white}{prismatic}};
  \node[anchor=south] at (0.44,-1.00) {\textcolor{white}{dislocation loop}};
  \draw[<-,white] (0.13,-0.66) -- (0.34,-0.71);
\end{tikzpicture}
\begin{tikzpicture}[x=1.000in,y=0.737in]
  \node[anchor=north west,inner sep=0] at (0,0) {
    \includegraphics[width=1in]{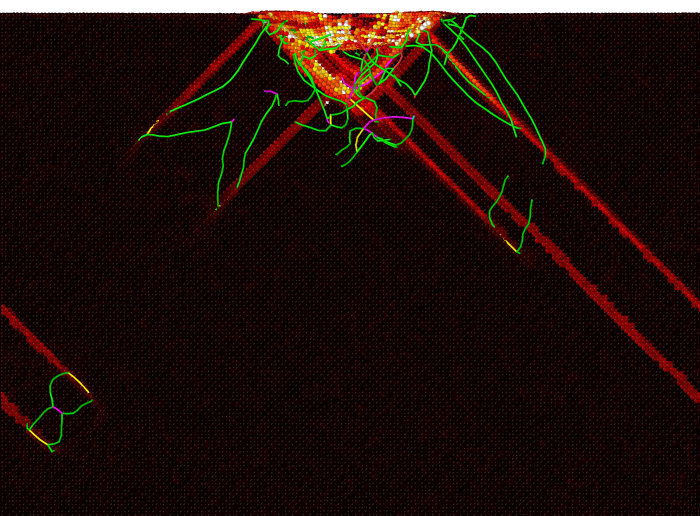}
  };
  \node[anchor=south] at (0.5,0.0) {ACE};
  \node[anchor=north west] at (0,0) {\textcolor{white}{c)}};
  \draw[<-,white] (0.3,-0.93) -- (0.75,-0.93) node[midway, above] {$[100]$};
  \draw[|-|] (1.05,0) -- (1.05,-1) node[midway, above, rotate=270] {$\SI{269.1}{\angstrom}$};
\end{tikzpicture}
\begin{tikzpicture}[x=1.000in,y=0.444in]
  \node[anchor=north west,inner sep=0] at (0,0) {
    \includegraphics[width=1in]{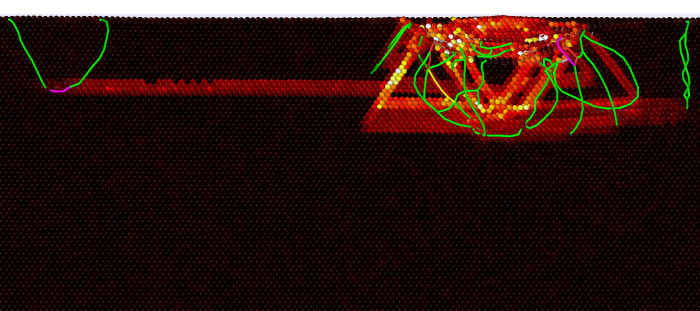}
  };
  \node[anchor=south west] at (0,-1) {\textcolor{white}{d)}};
  \node[anchor=south east] at (1,-0.85) {\textcolor{white}{wedge}};
  \node[anchor=south east] at (1,-1.00) {\textcolor{white}{dislocation}};
  \draw[<-,white] (0.1,-0.35) -- (0.38,-0.83);
\end{tikzpicture}
\begin{tikzpicture}[x=1.000in,y=0.444in]
  \node[anchor=north west,inner sep=0] at (0,0) {
    \includegraphics[width=1in]{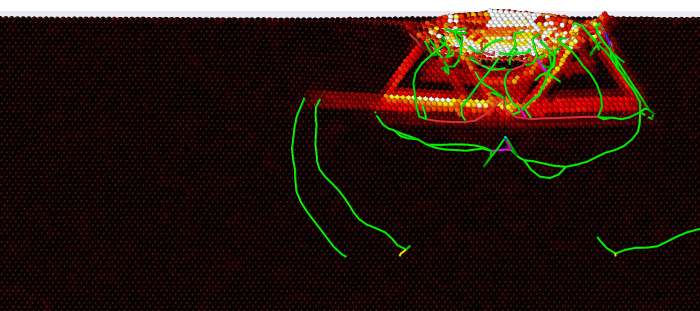}
  };
  \node[anchor=south west] at (0,-1) {\textcolor{white}{e)}};
\end{tikzpicture}
\begin{tikzpicture}[x=1.000in,y=0.444in]
  \node[anchor=north west,inner sep=0] at (0,0) {
    \includegraphics[width=1in]{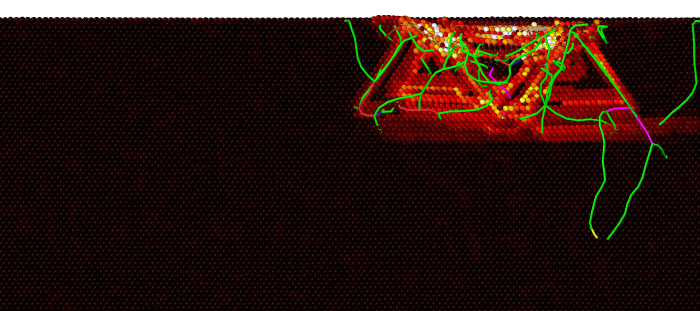}
  };
  \node[anchor=south west] at (0,-1) {\textcolor{white}{f)}};
  \draw[<-,white] (0.3,-0.88) -- (0.75,-0.88) node[midway, above] {$[\overline{1}\overline{1}0]$};
  \draw[|-|] (1.05,0) -- (1.05,-1) node[midway, above, rotate=270] {$\SI{140}{\angstrom}$};
\end{tikzpicture}
\begin{tikzpicture}[x=1.000in,y=0.050in]
  \node[anchor=north west,inner sep=0] at (0,0) {
    \frame{\includegraphics[width=1in]{Tungsten_plots_prod_shear_strain_color_bar.png}}
  };
  \node[anchor=south] at (0.5,0.0) {shear strain};
  \node[anchor=east] at (0.0,-0.5) {0};
  \node[anchor=west] at (1.0,-0.5) {1.5};
\end{tikzpicture}
\begin{tikzpicture}[x=3.300in,y=0.111in]
  \node[anchor=north west,inner sep=0] at (0,0) {
    \includegraphics[width=3.3in]{Copper_plots_prod_dislocation_type_legend.png}
  };
  \node[anchor=north] at (0.083,-1.0) {$1/6\langle112\rangle$};
  \node[anchor=north] at (0.250,-1.0) {$1/6\langle110\rangle$};
  \node[anchor=north] at (0.417,-1.0) {$1/3\langle100\rangle$};
  \node[anchor=north] at (0.583,-1.0) {$1/2\langle110\rangle$};
  \node[anchor=north] at (0.750,-1.0) {$1/3\langle111\rangle$};
  \node[anchor=north] at (0.916,-1.0) {other};
  \node[anchor=south] at (0.5,0.0) {dislocation type};
\end{tikzpicture}
\caption{\label{Fig::shear_strain_copper}
The copper atoms are color-coded according to the shear strain evaluated using OVITO in the a-c) $(010)$ and d-f) $(1\overline{1}0)$ plane through the contact point for a normalized indentation depth of $h/a_\text{c}\approx0.45$.
The dislocation lines computed with the DXA analysis of OVITO are shown for $\SI{35}{\angstrom}$ in the normal direction of the corresponding plane and color-coded according to their dislocation type.
A scale bar is given on the right side for every row.
}
\end{figure}

The shear strain was evaluated using OVITO according to the method described in Ref. \cite{2007MJ200769}, where the initial configuration was used as a reference. The shear strain is visualized at the end of the simulations in \cref{Fig::shear_strain_copper}.
The red slip traces in \cref{Fig::shear_strain_copper}(a-c) in the $[10\overline{1}]$ and $[\overline{1}0\overline{1}]$ directions with a shear strain of about 0.3 mark the glide of prismatic dislocation loops.
Furthermore, a wedge-shaped dislocation develops close the surface in the EAM simulation (see \cref{Fig::shear_strain_copper}d) like in Ref. \cite{HUANG2021110237} and glides in the $[\overline{1}\overline{1}0]$ direction.
This leads $1/2[\overline{1}\overline{1}0]$ atomic displacements which are visible at the surface in \cref{Fig::displacement_surface_copper}a.

\subsection{Tungsten}
\label{sec::results_tungsten}
\begin{table}
\caption{\label{tab::setup_parameters_tungsten}
Parameters and the used values in the tungsten nanoindentations performed with molecular dynamics simulations.
}
\centering
\begin{tabular}{lr}
\hline\hline
parameter & value\\\hline
surface orientation & (100)\\
temperature & $\SI{300}{\kelvin}$\\
initial box size & $\SI{273.1}{\angstrom} \times \SI{273.1}{\angstrom} \times \SI{271.9}{\angstrom}$\\
number of atoms & $1,272,112$\\
height fixed region & $\SI{20}{\angstrom}$\\
height NVT region & $\SI{10}{\angstrom}$\\
timestep $\Delta t$ & $\SI{1}{\femto\second}$\\
indenter radius $R_\text{ind}$ & $\SI{60}{\angstrom}$\cite{tungsten_nanoindentation_different_potentials_roomtemperature}\\
indenter velocity $\vec{v}_\text{ind}$ & $\{0,0,20\}\si{\metre/\second}$\cite{tungsten_nanoindentation_different_potentials_roomtemperature}\\
indenter force constant $k_\text{ind}$ & $\SI{236}{\electronvolt\angstrom^{-3}}$\cite{tungsten_nanoindentation_different_potentials_roomtemperature}\\
indentation depth $h_\text{max}$ & $\SI{20.8}{\angstrom}$\\
initial indentation depth $h_0$ & $\SI{-4.2}{\angstrom}$\\
\hline\hline
\end{tabular}
\end{table}

\begin{figure}[tb]
\includegraphics[width=3.37in]{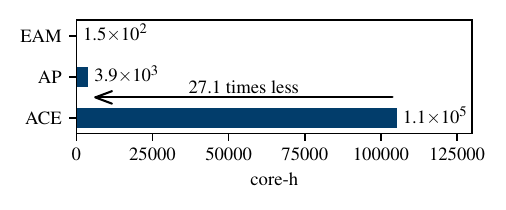}
\caption{\label{Fig::speedup_nanoindentation_tungsten}
Total computation time of a nanoindentation with 1.27 million W atoms simulated for $\SI{125}{\pico\second}$ using the AP potential compared to ACE and EAM simulations. EAM, AP and ACE simulations are calculated on 128, 128 and 640 cores of JURECA-DC\cite{jureca}, respectively.
}
\end{figure}

The nanoindentation simulation for W was carried out in equivalent manner as that for Cu. The parameters of the simulation are given in \cref{tab::setup_parameters_tungsten}. The speedup of the AP simulation compared to the ACE simulation is 27.1, as shown in \cref{Fig::speedup_nanoindentation_tungsten}.
Details regarding the computational efficiency are provided in \cref{sec::computational:efficiency}.

\begin{figure}[tb]
\begin{center}
\begin{tikzpicture}[x=1.1in,y=1.1in]
  \node[anchor=north west,inner sep=0] at (0,0) {
    \includegraphics[width=1.1in]{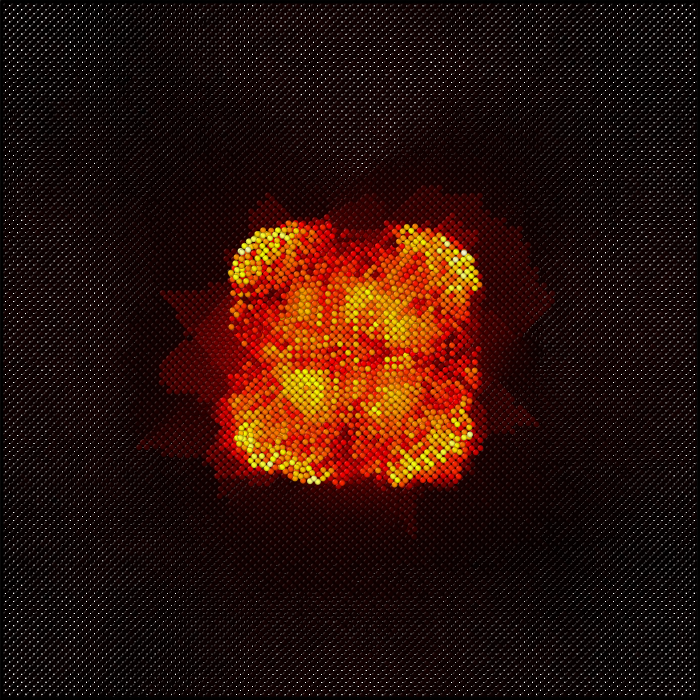}
  };
  \node[anchor=north] at (0.5,0.0) {\textcolor{white}{EAM}};
  \node[anchor=north west] at (0,0) {\textcolor{white}{a)}};
  \node [coordinate] (tripod_origin) at (0.15,-0.90) [] {};
  \draw[white,->] (tripod_origin) -- ($(tripod_origin) + (0.1,0)$) node[anchor=west]  {$[100]$};
  \draw[white,->] (tripod_origin) -- ($(tripod_origin) + (0,0.1)$) node[anchor=south] {$[010]$};
\end{tikzpicture}
\begin{tikzpicture}[x=1.1in,y=1.1in]
  \node[anchor=north west,inner sep=0] at (0,0) {
    \includegraphics[width=1.1in]{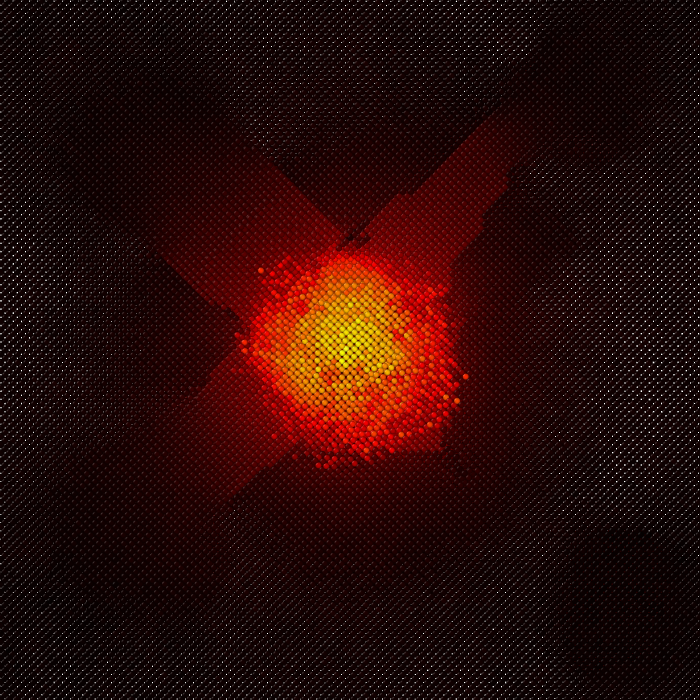}
  };
  \node[anchor=north] at (0.5,0.0) {\textcolor{white}{AP}};
  \node[anchor=north west] at (0,0) {\textcolor{white}{b)}};
\end{tikzpicture}
\begin{tikzpicture}[x=1.1in,y=1.1in]
  \node[anchor=north west,inner sep=0] at (0,0) {
    \includegraphics[width=1.1in]{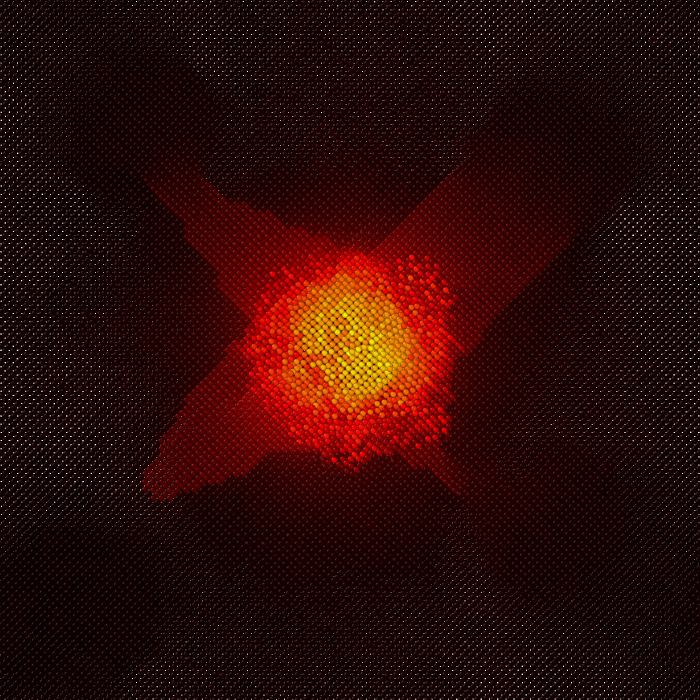}
  };
  \node[anchor=north] at (0.5,0.0) {\textcolor{white}{ACE}};
  \node[anchor=north west] at (0,0) {\textcolor{white}{c)}};
\end{tikzpicture}
\begin{tikzpicture}[x=1.000in,y=0.050in]
  \node[anchor=north west,inner sep=0] at (0,0) {
    \frame{\includegraphics[width=1in]{Tungsten_plots_prod_shear_strain_color_bar.png}}
  };
  \node[anchor=south] at (0.5,0.0) {$\Delta r / \si{\angstrom}$};
  \node[anchor=east] at (0.0,-0.5) {0};
  \node[anchor=west] at (1.0,-0.5) {30};
\end{tikzpicture}
 \includegraphics[width=\columnwidth]{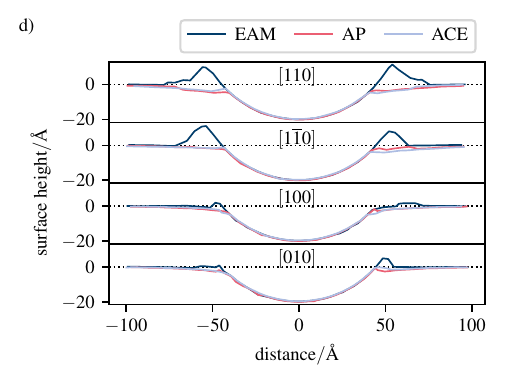}
\end{center}
\caption{\label{Fig::displacement_surface_tungsten}
a-c) Displacement $\Delta r$ of the W atoms at $\SI{20.8}{\angstrom}$ indentation depth compared with $\SI{-4.2}{\angstrom}$ indentation depth.
d) Cross sections through the center of the supercell below the indenter along different directions at $\SI{20.8}{\angstrom}$ indentation depth.}
\end{figure}

The displacement of the surface atoms between initial and maximum indentation depth is visualized in \cref{Fig::displacement_surface_tungsten}(a-c). In contrast to Cu, the observed surface pattern from the EAM simulation differs markedly from those of the AP and ACE simulations. For EAM, marked pile-ups are observed along the $\langle 110 \rangle$ directions while for AP and ACE the surfaces around the indenter remain almost flat. The variations of the surface height are plotted for different cross sections through the contact point of surface and indenter in \cref{Fig::displacement_surface_tungsten}(d).
EAM shows a pile-up of about $\SI{10.2\pm 1.3}{\angstrom}$ in the $\langle 110 \rangle$ directions while small depressions develop in the AP and ACE simulations.

Our results can be compared with those of other nanoindentation simulations \cite{tungsten_nanoindentation_different_potentials_roomtemperature, tungsten_nanoindentation_different_potentials_roomtemperature_arxiv} that used the same rigid spherical indenter and the same indentation velocity. In these studies, several interatomic potentials, including a tabulated Gaussian approximation potential (tabGAP) \cite{PhysRevMaterials.6.083801}, were systematically compared.  The tabGAP potential is an ML potential \cite{gaussian_approximation_potentials} which contains up to three-body descriptors and was extensively tested for various defects in W~\cite{PhysRevMaterials.6.083801}.
Similar to our study, pile-ups were found for the EAM potential while surface depressions were predicted by tabGAP at $\SI{30}{\angstrom}$ indentation depth \cite{tungsten_nanoindentation_different_potentials_roomtemperature_arxiv}.

\begin{figure}[tb]
\begin{center}
 \includegraphics[width=\columnwidth]{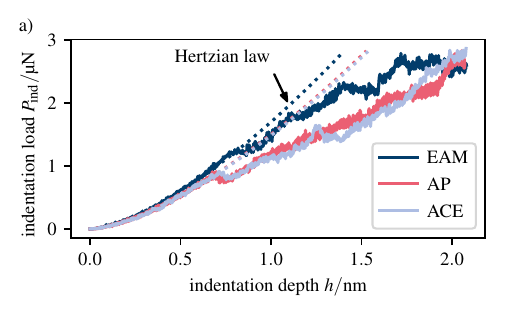}
 \includegraphics[width=\columnwidth]{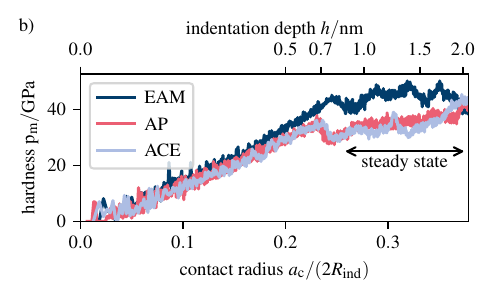}
\end{center}
\caption{\label{Fig::hertz_fit_tungsten}
a) Indentation load $P_\text{ind}$ and elastic Hertzian load $P_\text{H}$ according to \cref{eq::P_ind,eq::hertz_fit} computed for the W nanoindentations.
b) Mean contact pressure or hardness $p_\text{m}$ according to \cref{eq::p_m} dependent on the contact radius $a_\text{c}$ according to \cref{eq::contact_radius}.
The contact radius is normalized by the radius $R_\text{ind}$ of the spherical indenter.
}
\end{figure}

The Hertzian loads $P_\text{H}$ fitted to the indentation loads $P_\text{ind}$ according to \cref{eq::hertz_fit} are plotted in \cref{Fig::hertz_fit_tungsten}(a). The obtained indentation moduli of $\SI{519.0}{\giga\pascal}$ for EAM, $\SI{458.2}{\giga\pascal}$ for AP and $\SI{452.9}{\giga\pascal}$ for ACE confirm the consistency of AP and ACE simulations. 

The hardness calculated according to \cref{eq::p_m} is shown in \cref{Fig::hertz_fit_tungsten}(b). The sample deforms elastically until $h\approx\SI{8.3}{\angstrom}$, $\SI{7.0}{\angstrom}$ and $\SI{7.4}{\angstrom}$ for EAM, AP and ACE simulations, respectively. The hardness values in the steady state of plastic deformation are $\SI{44.3\pm2.6}{\giga\pascal}$ for EAM, $\SI{35.6\pm3.0}{\giga\pascal}$ for the AP potential and $\SI{35.1\pm4.1}{\giga\pascal}$ for ACE.
The pile-up, which develops in the EAM simulation, increases the contact radius while the surface depression in the AP and ACE simulations reduces the contact radius. Neither of these effects is incorporated in the analytical model of the contact radius in \cref{eq::contact_radius}. Thus, the hardness is overestimated in the EAM simulation compared to the AP and ACE simulations, which is consistent with the observed values. Our hardness values are comparable with the value of  $\SI{37.8}{\giga\pascal}$ reported in Ref. \cite{tungsten_nanoindentation_eam_all_temperatures} for another EAM potential~\cite{PhysRevB.69.144113}, where the indentation velocity of $\SI{50}{\metre/\second}$ and the force constant of $\SI{10}{\electronvolt\angstrom^{-3}}$ were applied. 

\begin{figure}[tb]
\begin{tikzpicture}[x=1.000in,y=0.5642857142857143in]
  \node[anchor=north west,inner sep=0] at (0,0) {
    \includegraphics[width=1in]{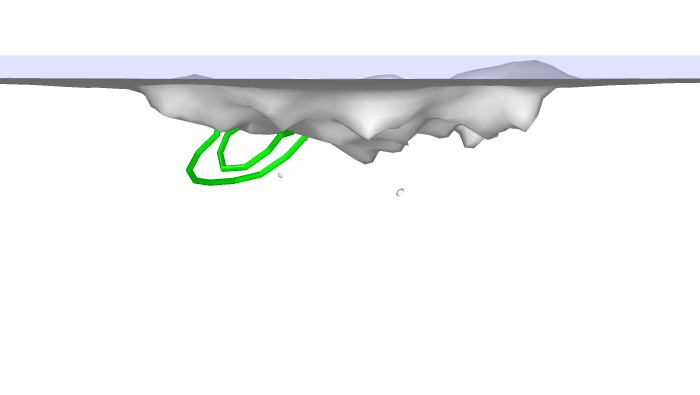}
  };
  \node[anchor=south] at (0.5,0.0) {EAM};
  \node[anchor=north west] at (0,-0.2) {\textcolor{black}{a)}};
  \node[anchor=south west] at (0, -1.0) {\textcolor{dxa_bcc_green}{\rule{0.2cm}{0.2cm}} $1/2\langle 111\rangle$};
\end{tikzpicture}
\begin{tikzpicture}[x=1.000in,y=0.5642857142857143in]
  \node[anchor=north west,inner sep=0] at (0,0) {
    \includegraphics[width=1in]{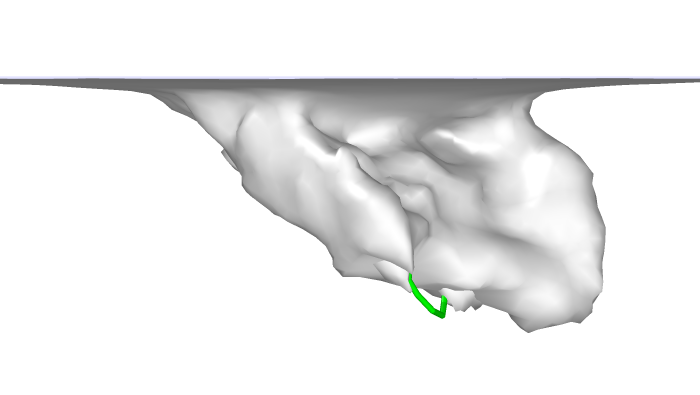}
  };
  \node[anchor=south] at (0.5,0.0) {AP};
  \node[anchor=north west] at (0,-0.2) {\textcolor{black}{b)}};
  \draw[->] (0.361,-0.390) -- (0.500,-0.603) node[midway, below, rotate=320] {$[\overline{1}1\overline{1}]$};
\end{tikzpicture}
\begin{tikzpicture}[x=1.000in,y=0.5642857142857143in]
  \node[anchor=north west,inner sep=0] at (0,0) {
    \includegraphics[width=1in]{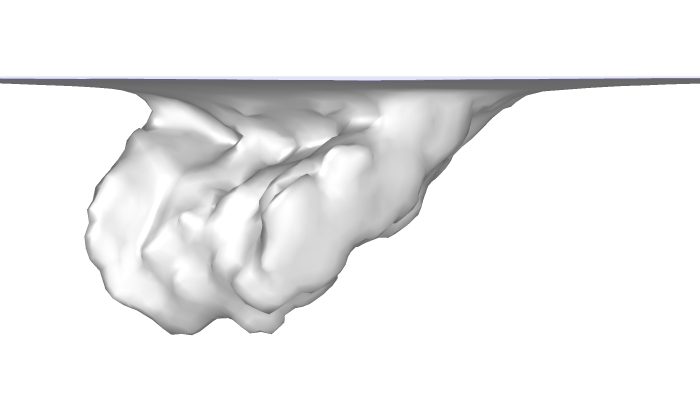}
  };
  \node[anchor=south] at (0.5,0.0) {ACE};
  \node[anchor=north west] at (0,-0.2) {\textcolor{black}{c)}};
  \draw[|-|] (1.05,0) -- (1.05,-1) node[midway, above, rotate=270] {$\SI{54.0}{\angstrom}$};
  \node [coordinate] (tripod_origin) at (0.85,-0.80) [] {};
  \draw[->] (tripod_origin) -- ($(tripod_origin) - (0.1,0)$) node[anchor=east]  {$[100]$};
  \draw[->] (tripod_origin) -- ($(tripod_origin) + (0,0.166)$) node[anchor=south] {$[001]$};
\end{tikzpicture}
\begin{tikzpicture}[x=1.000in,y=0.5642857142857143in]
  \node[anchor=north west,inner sep=0] at (0,0) {
    \includegraphics[width=1in]{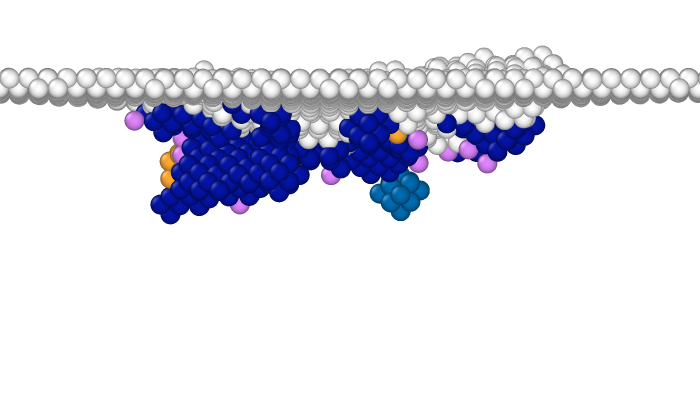}
  };
  \node[anchor=north west] at (0,-0.2) {\textcolor{black}{d)}};
\end{tikzpicture}
\begin{tikzpicture}[x=1.000in,y=0.5642857142857143in]
  \node[anchor=north west,inner sep=0] at (0,0) {
    \includegraphics[width=1in]{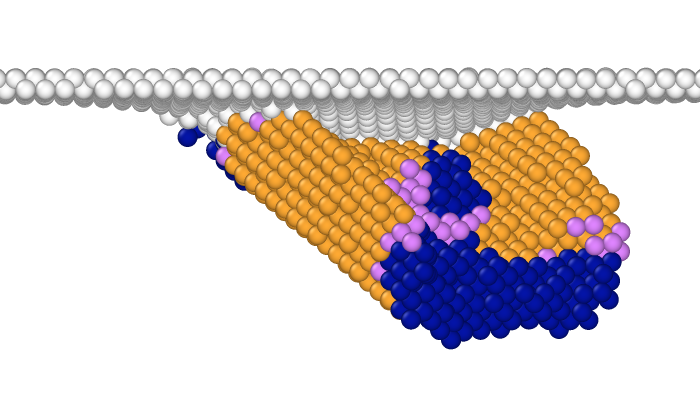}
  };
  \node[anchor=north west] at (0,-0.2) {\textcolor{black}{e)}};
  \draw[->] (0.361,-0.390) -- (0.500,-0.603) node[midway, below, rotate=320] {$[\overline{1}1\overline{1}]$};
\end{tikzpicture}
\begin{tikzpicture}[x=1.000in,y=0.5642857142857143in]
  \node[anchor=north west,inner sep=0] at (0,0) {
    \includegraphics[width=1in]{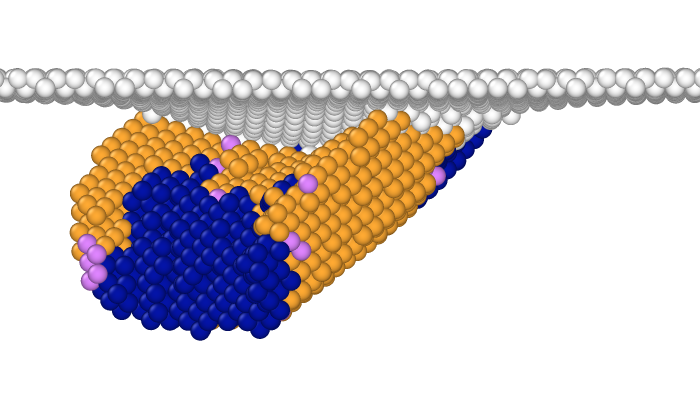}
  };
  \node[anchor=north west] at (0,-0.2) {\textcolor{black}{f)}};
  \draw[|-|] (1.05,0) -- (1.05,-1) node[midway, above, rotate=270] {$\SI{54.0}{\angstrom}$};
  \node [coordinate] (tripod_origin) at (0.85,-0.80) [] {};
  \draw[->] (tripod_origin) -- ($(tripod_origin) - (0.1,0)$) node[anchor=east]  {$[100]$};
  \draw[->] (tripod_origin) -- ($(tripod_origin) + (0,0.166)$) node[anchor=south] {$[001]$};
\end{tikzpicture}
\begin{tikzpicture}[x=1.000in,y=0.852857142857143in]
  \node[anchor=north west,inner sep=0] at (0,0) {
    \includegraphics[width=1in]{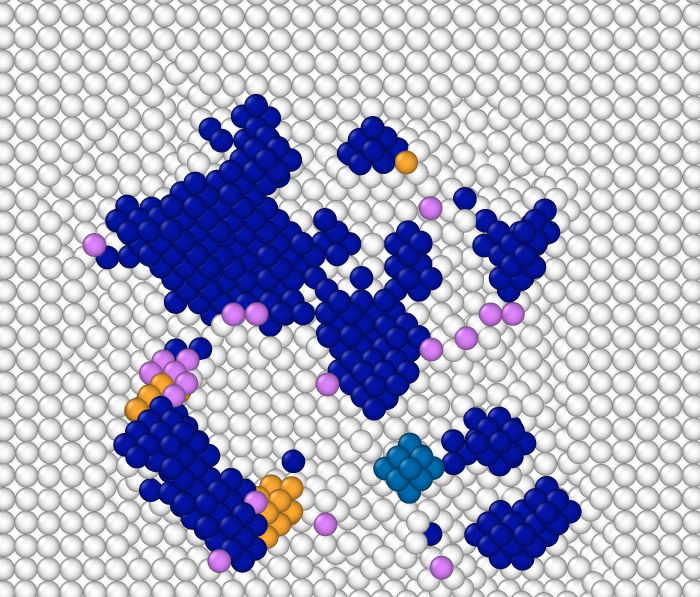}
  };
  \node[anchor=north west] at (0,0) {\textcolor{black}{g)}};
\end{tikzpicture}
\begin{tikzpicture}[x=1.000in,y=0.852857142857143in]
  \node[anchor=north west,inner sep=0] at (0,0) {
    \includegraphics[width=1in]{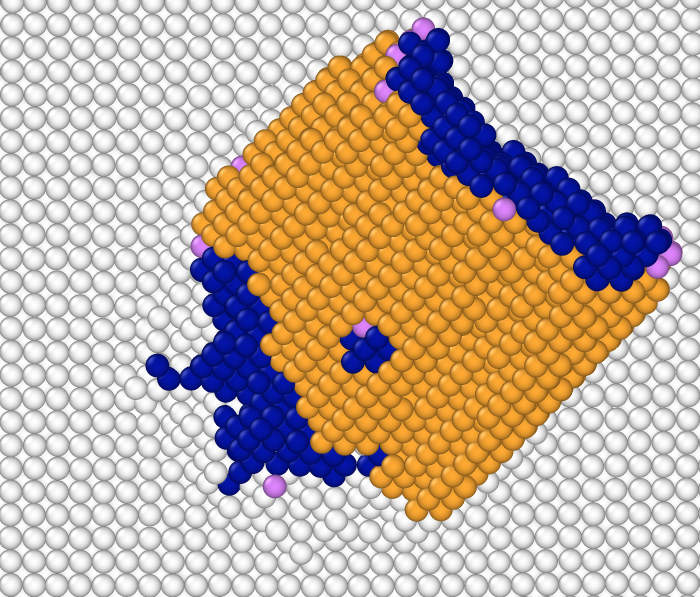}
  };
  \node[anchor=north west] at (0,0) {\textcolor{black}{h)}};
\end{tikzpicture}
\begin{tikzpicture}[x=1.000in,y=0.852857142857143in]
  \node[anchor=north west,inner sep=0] at (0,0) {
    \includegraphics[width=1in]{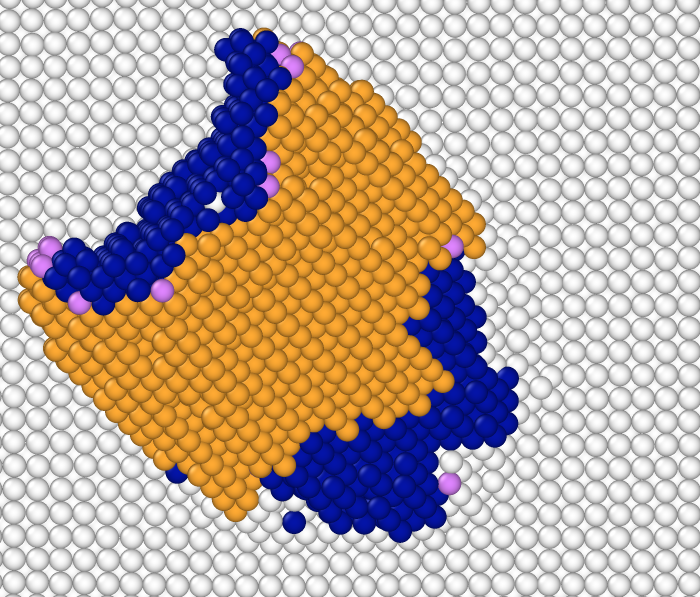}
  };
  \node[anchor=north west] at (0,0) {\textcolor{black}{i)}};
  \draw[|-|] (1.05,0) -- (1.05,-1) node[midway, above, rotate=270] {$\SI{81.5}{\angstrom}$};
  \node [coordinate] (tripod_origin) at (0.85,-0.15) [] {};
  \draw[->] (tripod_origin) -- ($(tripod_origin) - (0.1,0)$) node[anchor=east]  {$[100]$};
  \draw[->] (tripod_origin) -- ($(tripod_origin) - (0,0.105)$) node[anchor=north] {$[0\overline{1}0]$};
\end{tikzpicture}
\begin{tikzpicture}[x=1.000in,y=0.852857142857143in]
  \draw[black,fill=bda_white]     (0.0,-0.0) circle (0.10cm) node[anchor=west] {\ surface};
  \draw[black,fill=bda_lightblue] (0.0,-0.2) circle (0.10cm) node[anchor=west] {\ vacancy};
  \draw[black,fill=bda_darkblue]  (0.7,-0.0) circle (0.10cm) node[anchor=west] {\ non-screw dislocation};
  \draw[black,fill=bda_orange]    (0.7,-0.2) circle (0.10cm) node[anchor=west] {\ twin/screw dislocation};
  \draw[black,fill=bda_green]     (2.1,-0.0) circle (0.10cm) node[anchor=west] {\ \{110\} planar fault};
  \draw[black,fill=bda_purple]    (2.1,-0.2) circle (0.10cm) node[anchor=west] {\ unidentified};
\end{tikzpicture}
\caption{\label{Fig::bda_networks_tungsten}
Visualization of dislocation lines using the DXA algorithm of OVITO (a-c) and BCC defects identified by BDA (d-i) for a normalized indentation depth of $h/a_\text{c}\approx 0.41$. In BDA, the atoms are color-coded according to the identified defect type while perfect bulk atoms are not shown.
A scale bar is given on the right side for every row.
}
\end{figure}
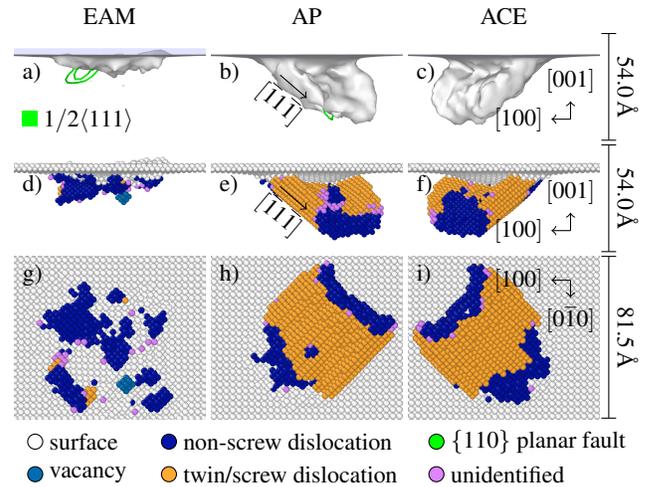

\begin{figure}[tb]
\begin{tikzpicture}[x=1.000in,y=0.601in]
  \node[anchor=north west,inner sep=0] at (0,0) {
    \includegraphics[width=1in]{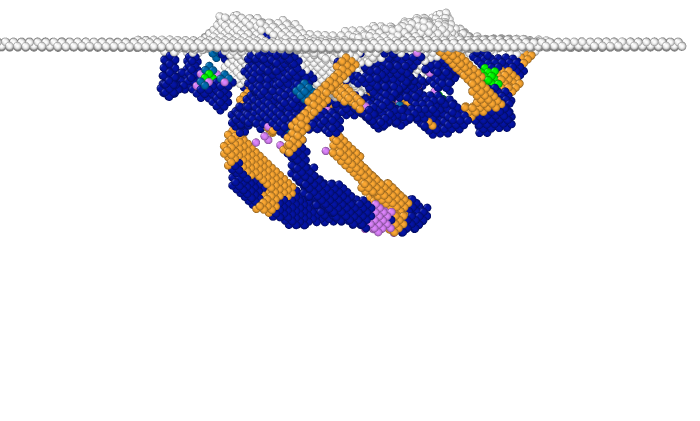}
  };
  \node[anchor=south] at (0.5,0.0) {EAM};
  \node[anchor=north west] at (0,-0.1) {\textcolor{black}{a)}};
  \node [coordinate] (tripod_origin) at (0.85,-0.85) [] {};
  \draw[->] (tripod_origin) -- ($(tripod_origin) - (0.1,0)$) node[anchor=east]  {$[100]$};
  \draw[->] (tripod_origin) -- ($(tripod_origin) + (0,0.166)$) node[anchor=south] {$[001]$};
\end{tikzpicture}
\begin{tikzpicture}[x=1.000in,y=0.601in]
  \node[anchor=north west,inner sep=0] at (0,0) {
    \includegraphics[width=1in]{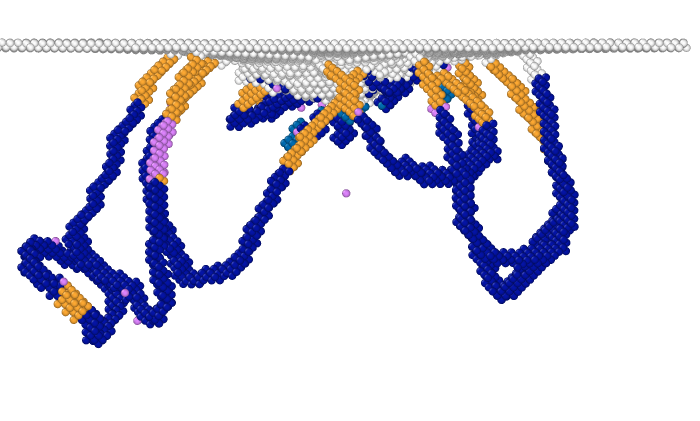}
  };
  \node[anchor=south] at (0.5,0.0) {AP};
  \node[anchor=north west] at (0,-0.1) {\textcolor{black}{b)}};
\end{tikzpicture}
\begin{tikzpicture}[x=1.000in,y=0.601in]
  \node[anchor=north west,inner sep=0] at (0,0) {
    \includegraphics[width=1in]{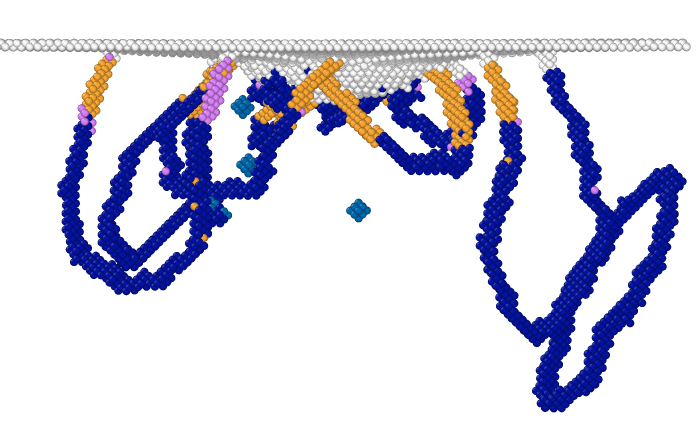}
  };
  \node[anchor=south] at (0.5,0.0) {ACE};
  \node[anchor=north west] at (0,-0.1) {\textcolor{black}{c)}};
  \draw[|-|] (1.05,0) -- (1.05,-1) node[midway, above, rotate=270] {$\SI{155.5}{\angstrom}$};
\end{tikzpicture}
\begin{tikzpicture}[x=1.000in,y=0.601in]
  \node[anchor=north west,inner sep=0] at (0,0) {
    \includegraphics[width=1in]{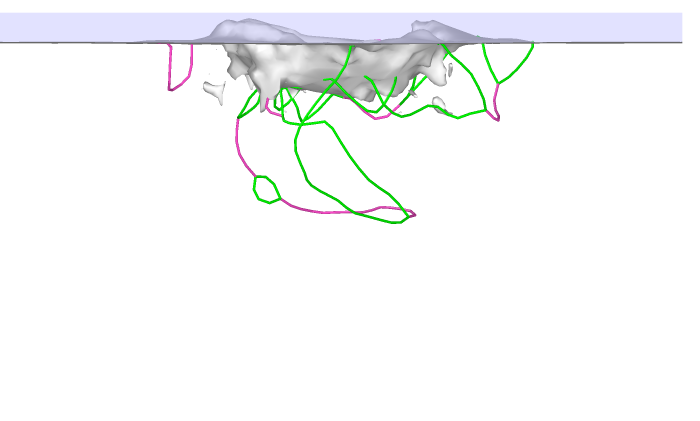}
  };
  \node[anchor=north west] at (0,-0.1) {\textcolor{black}{d)}};
  \node [coordinate] (tripod_origin) at (0.85,-0.85) [] {};
  \draw[->] (tripod_origin) -- ($(tripod_origin) - (0.1,0)$) node[anchor=east]  {$[100]$};
  \draw[->] (tripod_origin) -- ($(tripod_origin) + (0,0.166)$) node[anchor=south] {$[001]$};
\end{tikzpicture}
\begin{tikzpicture}[x=1.000in,y=0.601in]
  \node[anchor=north west,inner sep=0] at (0,0) {
    \includegraphics[width=1in]{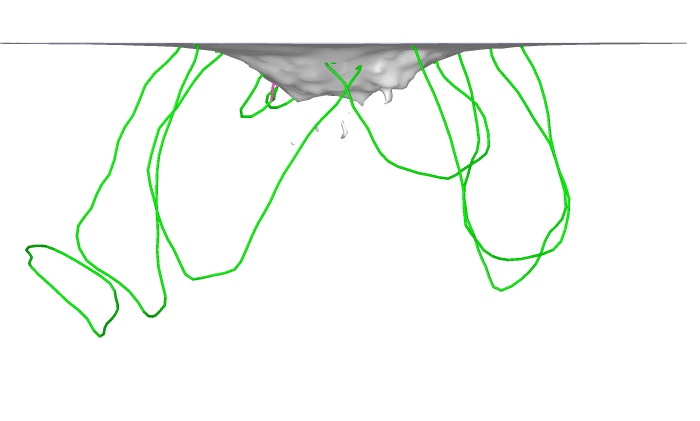}
  };
  \node[anchor=north west] at (0,-0.1) {\textcolor{black}{e)}};
\end{tikzpicture}
\begin{tikzpicture}[x=1.000in,y=0.601in]
  \node[anchor=north west,inner sep=0] at (0,0) {
    \includegraphics[width=1in]{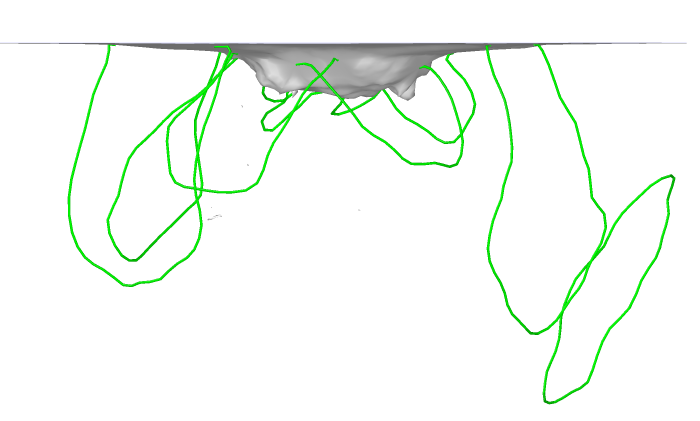}
  };
  \node[anchor=north west] at (0,-0.1) {\textcolor{black}{f)}};
  \draw[|-|] (1.05,0) -- (1.05,-1) node[midway, above, rotate=270] {$\SI{155.5}{\angstrom}$};
\end{tikzpicture}
\begin{tikzpicture}[x=1.000in,y=0.957in]
  \node[anchor=north west,inner sep=0] at (0,0) {
    \includegraphics[width=1in]{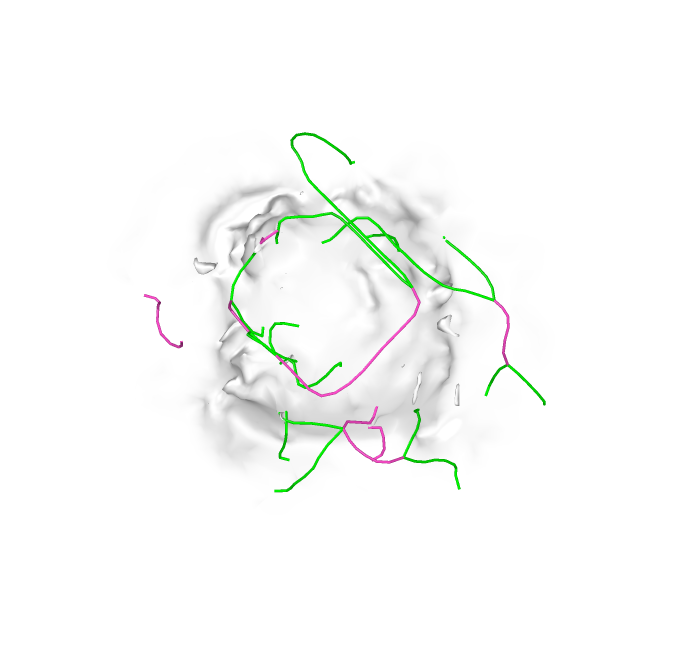}
  };
  \node[anchor=north west] at (0,-0.3) {\textcolor{black}{g)}};
  \draw[dotted,fzjred] (0.021323945702308666,-0.9517446988501097) -- (0.7891365923209812,-0.150267960049962);
  \node [coordinate] (tripod_origin) at (0.85,-0.92) [] {};
  \draw[->] (tripod_origin) -- ($(tripod_origin) - (0.1,0)$) node[anchor=east]  {$[100]$};
  \draw[->] (tripod_origin) -- ($(tripod_origin) + (0,0.105)$) node[anchor=south] {$[010]$};
  \node[anchor=north west,inner sep=0] at (0,-1.1) {
    \frame{\includegraphics[width=1in]{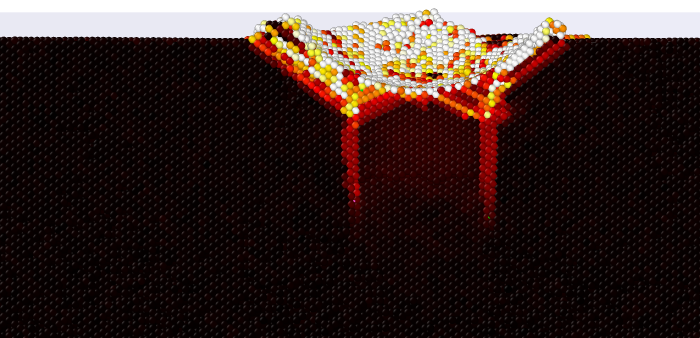}}
  };
  \node[anchor=south west] at (0,-1.6) {\textcolor{white}{j)}};
  \draw[<-,white] (0.4,-1.56) -- (0.95,-1.56) node[midway, above] {$[1\overline{1}0]$};
\end{tikzpicture}
\begin{tikzpicture}[x=1.000in,y=0.957in]
  \node[anchor=north west,inner sep=0] at (0,0) {
    \includegraphics[width=1in]{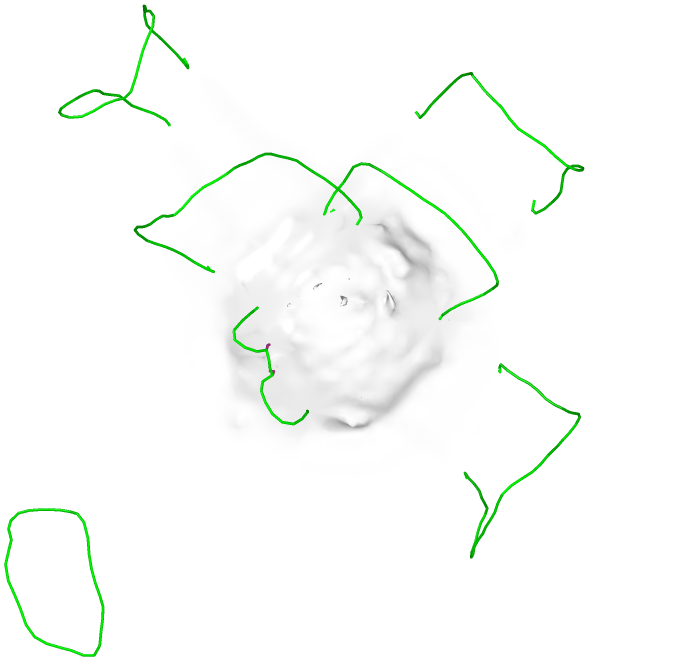}
  };
  \node[anchor=north west] at (0,-0.3) {\textcolor{black}{h)}};
  \draw[dotted, fzjred] (0.021323945702308666,-0.9517446988501097) -- (0.7891365923209812,-0.150267960049962);
  \node[anchor=north west,inner sep=0] at (0,-1.1) {
    \frame{\includegraphics[width=1in]{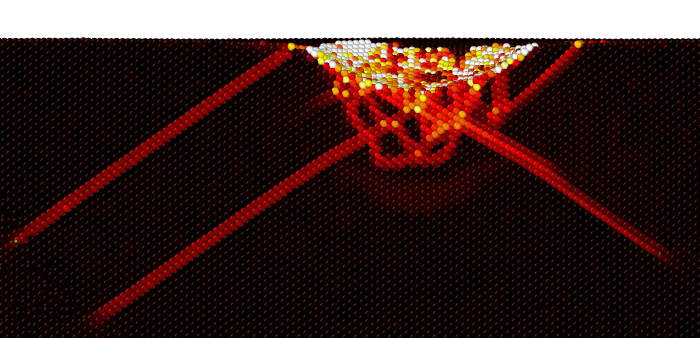}}
  };
  \draw[dashed, ->,fzjred] (0.021323945702308666,-0.9517446988501097) -- (0,-1.1);
  \draw[dashed, ->,fzjred] (0.7891365923209812,-0.150267960049962) -- (1,-1.1);
  \node[anchor=south west] at (0,-1.6) {\textcolor{white}{k)}};
\end{tikzpicture}
\begin{tikzpicture}[x=1.000in,y=0.957in]
  \node[anchor=north west,inner sep=0] at (0,0) {
    \includegraphics[width=1in]{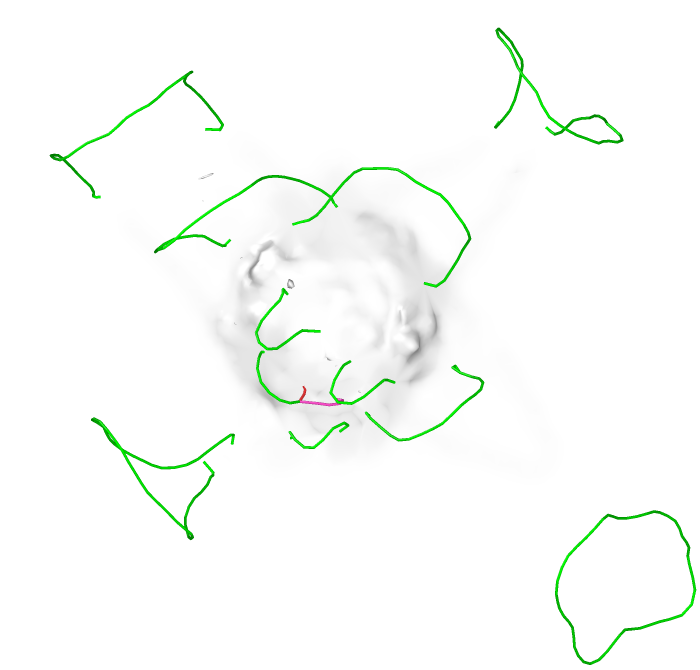}
  };
  \node[anchor=north west] at (0,-0.3) {\textcolor{black}{i)}};
  \draw[dotted,fzjred] (0.021323945702308666,-0.9517446988501097) -- (0.7891365923209812,-0.150267960049962);
  \draw[|-|] (1.05,0) -- (1.05,-1) node[midway, above, rotate=270] {$\SI{247.6}{\angstrom}$};
  \node[anchor=north west,inner sep=0] at (0,-1.1) {
    \frame{\includegraphics[width=1in]{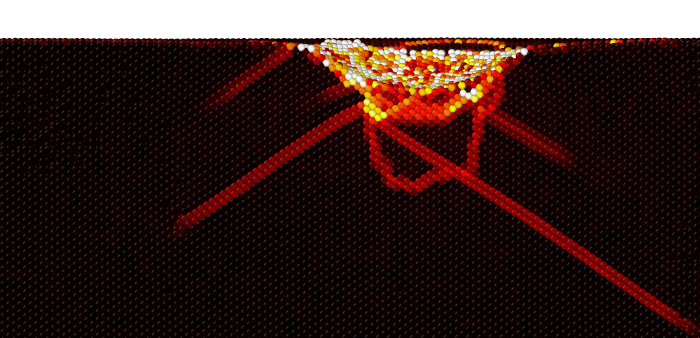}}
  };
  \node[anchor=south west] at (0,-1.6) {\textcolor{white}{l)}};
  \draw[|-|] (1.05,-1.1) -- (1.05,-1.603657262277952) node[midway, above, rotate=270] {$\SI{135.8}{\angstrom}$};

\end{tikzpicture}
\begin{tikzpicture}[x=1.6in,y=0.0376in]
  \node[anchor=north west,inner sep=0] at (0,0) {
    \includegraphics[width=1.6in]{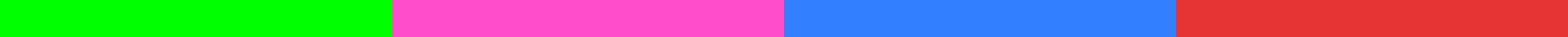}
  };
  \node[anchor=north] at (0.125,-1.0) {$1/2\langle111\rangle$};
  \node[anchor=north] at (0.375,-1.0) {$\langle100\rangle$};
  \node[anchor=north] at (0.625,-1.0) {$\langle110\rangle$};
  \node[anchor=north] at (0.875,-1.0) {other};
  \node[anchor=south] at (0.5,0.0) {dislocation type};
\end{tikzpicture}
\begin{tikzpicture}[x=1.000in,y=0.050in]
  \node[anchor=north west,inner sep=0] at (0,0) {
    \frame{\includegraphics[width=1in]{Tungsten_plots_prod_shear_strain_color_bar.png}}
  };
  \node[anchor=south] at (0.5,0.0) {shear strain};
  \node[anchor=east] at (0.0,-0.5) {0};
  \node[anchor=west] at (1.0,-0.5) {1.5};
\end{tikzpicture}
\begin{tikzpicture}[x=1.000in,y=0.852857142857143in]
  \draw[black,fill=bda_white]     (0.0,-0.0) circle (0.10cm) node[anchor=west] {\ surface};
  \draw[black,fill=bda_lightblue] (0.0,-0.2) circle (0.10cm) node[anchor=west] {\ vacancy};
  \draw[black,fill=bda_darkblue]  (0.7,-0.0) circle (0.10cm) node[anchor=west] {\ non-screw dislocation};
  \draw[black,fill=bda_orange]    (0.7,-0.2) circle (0.10cm) node[anchor=west] {\ twin/screw dislocation};
  \draw[black,fill=bda_green]     (2.1,-0.0) circle (0.10cm) node[anchor=west] {\ \{110\} planar fault};
  \draw[black,fill=bda_purple]    (2.1,-0.2) circle (0.10cm) node[anchor=west] {\ unidentified};
\end{tikzpicture}
\caption{\label{Fig::dislocation_networks_tungsten}
a-c) Visualisation of the atoms detected by the BCC defect analysis for the maximum normalised indentation depth of $h/a_\text{c}\approx 0.62$.
d-i) Visualisation of the dislocation lines identified by the dislocation extraction algorithm viewed from different perspectives. The dotted line visualises the (110) plane.
j-l) A cut in this plane through the system is used to visualise the by OVITO calculated shear strain.
Scale bars are given on the right side of each row.
The visualisations in the first column are from the EAM simulation, the second column contains the ones of the adaptive-precision (AP) simulation and the ACE simulation is shown in the third column.
}
\end{figure}

A convenient tool to identify defects in BCC materials is the BCC defect analysis (BDA) \cite{bda}, which has been employed for inspection of several nanoindentation simulations \cite{tungsten_nanoindentation_different_potentials_roomtemperature,NAGHDI2024120200}.
The BDA identification is based on common neighbor analysis \cite{cna}, coordination number and centro-symmetry parameter.
The algorithm can identify surface atoms, atoms next to a vacancy, screw and non-screw dislocations, \{110\} planar faults and twin boundaries. The screw dislocations and twin boundaries cannot be distinguished by BDA, but the line character of screw dislocations and planar character of twin boundaries can be discerned easily by a visual inspection. Since deformation twinning may compete with dislocation-mediated plasticity in W nanocrystals~\cite{wang2015situ}, BDA provides a complementary view to the detection of dislocation lines with DXA. Both DXA and BDA visualizations for the formalized indentation depth $h/a_\text{c}\approx{0.41}$ are shown in \cref{Fig::bda_networks_tungsten}. Furthermore, the DXA and BDA results for the maximum indentation depth $h/a_\text{c}\approx{0.62}$ are shown in \cref{Fig::dislocation_networks_tungsten}.

For EAM, two $1/2\langle 111\rangle$ dislocation loops are visible in \cref{Fig::bda_networks_tungsten}(a), as they emerge from the highly distorted region below the indenter. In contrast, a deformation twin develops initially in both AP and ACE simulations. The twin structures grow in the $\langle 111 \rangle$ directions on the $\{211\}$ habit planes. Eventually, the twin propagation stops and $1/2\langle 111\rangle$ dislocation loops start to nucleate from the twin tip, as visible in \cref{Fig::bda_networks_tungsten}(b). This process likely occurs by coalescence of three $1/6\langle 111\rangle$ twinning dislocations and leads to gradual retraction of the twin and transformation of the twinned region into a dislocation network. At the maximum indentation depth, as shown in \cref{Fig::dislocation_networks_tungsten}(b,c), only dislocation loops are present below the indenter. Such incipient nanocontact plasticity via nucleation, propagation and annihilation of twins has been reported in simulations of BCC Ta~\cite{Alcala_PhysRevLett.109.075502}. A similar outcome was also observed in the recent tabGAP simulations \cite{tungsten_nanoindentation_different_potentials_roomtemperature}, but some twins were still present at the maximum indentation depth of $\SI{30}{\angstrom}$ while all twins transformed into dislocation lines in our simulations. The initial twinning deformation was not observed for any other classical potentials used in Ref. \cite{tungsten_nanoindentation_different_potentials_roomtemperature}.

Dislocation lines identified by the DXA and BDA approaches are shown in \cref{Fig::dislocation_networks_tungsten} for the maximum indentation depth of $h/a_\text{c}\approx 0.62$. While DXA can distinguish BCC dislocations with the $1/2\langle 111\rangle$, $\langle 100\rangle$ and $\langle 110\rangle$ Burgers vectors, BDA can resolve only the $1/2\langle 111\rangle$ screw and non-screw types.  The dislocation network in EAM simulations differs markedly from those in AP and ACE simulations. In the latter simulations, one can see almost exclusively dislocations with the shortest $1/2\langle 111\rangle$ Burgers vector. The dislocation half-loops emanating from the indented region consist of long segments of predominantly screw character connected by a curved mixed/edge segment. The presence of long screws is expected due to their large Peierls stress and thermally-activated motion. The half-loops propagate away from the indenter and their interaction leads in some cases to formation of prismatic dislocation loops, as observed also in Ref. \cite{tungsten_nanoindentation_different_potentials_roomtemperature}. These loops continue to glide in the $\langle 111\rangle$ directions as evidenced by the slip lines in \cref{Fig::dislocation_networks_tungsten}(k,l).

In contrast, there are no extended dislocation segments or $\langle 111\rangle$ loops observed in the EAM simulation, as shown in \cref{Fig::dislocation_networks_tungsten}(d). Instead, the dislocation network is concentrated primarily right under the indenter with significant portion of $\langle 100\rangle$ dislocations that glide in the $[00\overline{1}]$ direction normal to the surface, as visible on the slip traces in \cref{Fig::dislocation_networks_tungsten}(j). This observation is consistent with results from Ref. \cite{tungsten_nanoindentation_different_potentials_roomtemperature}, where $\langle 100\rangle$ shear loops, which nucleated below the indenter and glided in $[00\overline{1}]$ direction, were observed in simulations that used the same EAM potential and indenter but a larger indentation depth.

\begin{figure}[tb]
\begin{center}
 \includegraphics[width=\columnwidth]{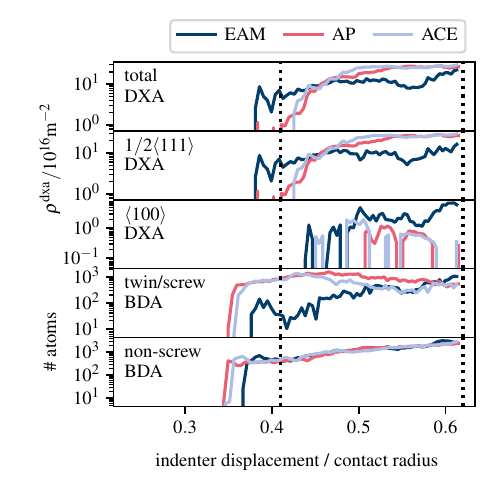}
\end{center}
\caption{\label{Fig::dislocation_density_tungsten}
Dislocation density evaluated using DXA and BDA analyses for nanoindentations of W$_{(100)}$. Since BDA cannot distinguish twins and screw dislocations, only their sum is given. The dashed vertical lines mark the indentation depths analyzed in \cref{Fig::dislocation_networks_tungsten,Fig::bda_networks_tungsten}.
}
\end{figure}

The dislocation density is calculated according to \cref{eq::dislocation_density}.
For this we need the radius $a_\text{pz}$ of the plastic zone according to \cref{eq::a_pz}, where the material-dependent factor $f_\text{pz}=1.9$ is used for W \cite{tungsten_nanoindentation_different_potentials_roomtemperature}.
The dislocation density as function of the indentation depth is shown in \cref{Fig::dislocation_density_tungsten}.
Furthermore, the plot includes also defect analysis using the BDA algorithm. The number of twins/screw dislocations has a  maximum at 0.47 and 0.43 for the AP and ACE simulations, respectively. This is consistent with the visual analysis discussed above, since twins gradually transform to $1/2\langle 111\rangle$ dislocations. Thus, \cref{Fig::dislocation_density_tungsten} confirms the transition from initial twin-mediated plasticity to dislocation-mediated plasticity in the AP and ACE simulations. For EAM, there is only dislocation-mediated plasticity.
Furthermore, one can clearly see the different mechanisms of dislocation-mediated plasticity between EAM simulation and ACE/AP simulations. The density of $\langle 100\rangle$ dislocations remains high during the course of the EAM simulation while these dislocations are only intermittent in the ACE/AP simulations.

\section{Conclusion}
\label{sec::conclusion}
In this work we systematically compared results of nanoindentation simulations for the prototypical FCC metal copper and BCC metal tungsten performed with interatomic potentials of different accuracy and computational cost. We employed computationally efficient embedded atom method (EAM) potentials, accurate but less efficient machine learning ACE potentials, and an AP combination of both \cite{adaptive_precision_potentials}. The selection of atoms treated by the AP potential was extended, compared to our previous work, to ensure an accurate description of atoms whose environment differs significantly from that of bulk crystal. 

Our results show that for Cu, all potentials yield similar dislocation morphologies under the indenter with only small quantitative differences. This confirms that the EAM potential can describe well the metallic bonding in free-electron metal Cu. Therefore, despite the achieved speedup of 21.1 of the AP simulation compared to the ACE simulation, one does not gain additional insights from the additional accuracy offered by the ACE potential in this case.

In contrast, markedly different plasticity mechanisms were observed for W in simulations performed with the central-force EAM potential compared to results obtained using the ACE potential, which is able to describe accurately the angular character of bonding in W due to its half-filled d-band. The EAM potential (also with optimized parameters as shown in \cref{sec::tungsten_nanoindentation_optimised_eam}) predicts dislocation-mediated plasticity and sustained presence of the $\langle 100\rangle$ dislocations below the indenter. In the ACE simulations, we observed instead a transition from initial twin-mediated plasticity to dislocation-mediated plasticity and predominance of the common $1/2\langle 111\rangle$ dislocations. All ACE-specific mechanisms were reproduced in the AP nanoindentation simulations, however, with a significant speedup of almost 30 times compared to the ACE-only simulations. Hence, the AP potential overcomes the performance gap between the precise ACE and the fast EAM potential by combining the advantages of both potentials and is beneficial for materials where simple central-force models are not appropriate.

\begin{acknowledgments}
We would like to thank A. Bochkarev for providing the preliminary ACE parametrization for tungsten.
The authors gratefully acknowledge computing time on the supercomputer JURECA\cite{jureca} at Forschungszentrum Jülich under grant no. 28990 (hybridace).
\end{acknowledgments}

\section*{Author declarations}
\subsection*{Competing interests}
The authors declare no competing interests.

\subsection*{Author Contributions}
\textbf{David Immel:}
Formal analysis (equal);
Investigation (equal);
Methodology (equal);
Resources (equal);
Software (lead);
Validation (equal);
Visualization (lead);
Writing - original draft (lead);
\textbf{Matous Mrovec:}
Formal analysis (equal);
Investigation (equal);
Methodology (equal);
Validation (equal);
Writing - original draft (supporting);
Writing - review \& editing (lead);
\textbf{Ralf Drautz:}
Conceptualization (equal);
Methodology (equal);
Supervision (equal);
Writing - review \& editing (supporting);
\textbf{Godehard Sutmann:}
Conceptualization (equal);
Methodology (equal);
Resources (equal);
Supervision (equal);
Writing - review \& editing (supporting);

\section*{Code availability}
Our modifications to the LAMMPS source code, that allow the usage of adaptive-precision interatomic potentials (APIP), are currently available at  \url{https://github.com/d-immel/lammps_ap/tree/apip}.\footnote{The staggered-grid domain decomposition for LAMMPS will be published separately in the future, but APIP does not depend on the geometry of the spatial domain decomposition\cite{spatial_domain_decomposition_lammps} and can be used also with the load-balancing via recursive coordinate bisectioning\cite{berger1987rcb}, that is implemented in LAMMPS\cite{lammps}.}
We plan that the APIP package will be available as part of LAMMPS under \url{https://github.com/lammps/lammps}
in the future.

\section*{Data availability}
The data that support the findings of this study are available from D.I. upon reasonable request.

\appendix

\section{Equilibration}
\label{sec::equilibration}
We use a relaxation technique which is adapted from Refs. \cite{relaxation_nanoindentation_originalsource,relaxation_nanoindentation_mysource}.
At first, a perfect lattice is created at $0\,\si{\kelvin}$.
The perfect lattice is simulated with periodic boundary conditions in all dimensions as NVT ensemble with a Langevin thermostat with the damping constant $\gamma^\text{L}$ until the temperature is within the tolerance $\delta_T$ to the target temperature $T_\text{targ}$.
In the second step, a NPT ensemble is simulated according to Nose-Hoover with damping parameters $\gamma^\text{NH}_T$ and $\gamma^\text{NH}_p$ for temperature and pressure until target temperature and target pressure $p_\text{targ}$ are within the tolerance $\delta_T$ and $\delta_p$ of the target values.
At this target temperature and pressure, the surface is created at the top of the simulation box in the $z$ direction and the open boundary conditions are used instead.
The bottom atoms in the $z$ direction are frozen at their current positions and their velocity is set to zero and not updated any more.
In a third simulation of the duration $\Delta t_\text{E}$ a NVT ensemble with a Langevin thermostat is applied.
The thermostat damps the pressure waves of the created surface.
The total momentum is set to zero repeatedly after the time interval $\Delta t_\text{mom}$ since it is not conserved by the Langevin thermostat.
All equilibration steps and the nanoindentation itself are simulated with the timestep $\Delta t$.

Dynamic load balancing is essential for the adaptive-precision simulations as the time required to calculate forces and energies for a particle heavily depends on the used interatomic potential.
We use a staggered grid\cite{rene_masterarbeit} as spatial domain-decomposition\cite{spatial_domain_decomposition_lammps} which is balanced by ALL\cite{ALL} as described in Ref. \cite{adaptive_precision_potentials}.
For the EAM and ACE simulation, we use the recursive coordinate bisectioning (RCB)\cite{berger1987rcb} from LAMMPS which balances LAMMPS' tiled domain layout according to the measured force-calculation time.
The load balancer is called after every time interval $\Delta t_\text{RCB}$ or $\Delta t_\text{SG}$ if the system is imbalanced enough.

All mentioned parameters are listed in \cref{tab::equilibration_parameters} for the copper and tungsten simulations with all three interatomic potentials.

\begin{table}[tp]
\caption{\label{tab::equilibration_parameters}
Parameters used during the three steps of equilibration and during the nanoindentation as described in \cref{sec::equilibration}.
}
\centering
\begin{tabular}{lrrrrrr}
\hline\hline
property                        & \multicolumn{3}{c}{copper} & \multicolumn{3}{c}{tungsten} \\
                                & EAM & AP & ACE             & EAM & AP & ACE \\
\hline
\multicolumn{7}{c}{step 1: NVT periodic boundaries} \\
$\gamma^\text{L} / \si{\pico\second}$ & 1 & 1 & 1 & 1 & 1 & 1 \\
$\delta_T / \si{\kelvin}$ & 0.2 & 0.2 & 0.2 & 0.2 & 0.2 & 0.2 \\
\multicolumn{7}{c}{step 2: NPT periodic boundaries} \\
$\delta_p / \si{\bar}$ & 10 & 10 & 10 & 10 & 10 & 10 \\
$\gamma^\text{NH}_T / \si{\pico\second}$ & 0.1 & 0.1 & 0.1 & 0.1 & 0.1 & 0.1 \\
$\gamma^\text{NH}_p / \si{\pico\second}$ & 1 & 1 & 1 & 1 & 1 & 1 \\
$p_\text{targ} / \si{\bar}$ & 0 & 0 & 0 &  0 & 0 & 0 \\
\multicolumn{7}{c}{step 3: NVT surface} \\
$\Delta t_\text{E} / \si{\pico\second}$ & 50 & 50 & 50 & 100 & 2 & 100 \\
$\gamma^\text{L} / \si{\pico\second}$ & 10 & 10 & 10 & 0.1 & 0.1 & 0.1 \\
$\Delta t_\text{mom} / \si{\pico\second}$ & 1 & 1 & 1 & 1 & 1 & 1 \\
\multicolumn{7}{c}{step 4: nanoindentation} \\
$\gamma^\text{NH}_T / \si{\pico\second}$ & 0.1 & 0.1 & 0.1 & 0.1 & 0.1 & 0.1 \\
\multicolumn{7}{c}{all steps} \\
$T_\text{targ} / \si{\kelvin}$ & 292 & 292 & 292 & 300 & 300 & 300 \\
$\Delta t / \si{\pico\second}$ & 0.001 & 0.001 & 0.001 & 0.001 & 0.001 & 0.001 \\
$\Delta t_\text{RCB} / \si{\pico\second}$ & 0.1 & - & 0.1 & 0.1 & - & 0.1 \\
$\Delta t_\text{SG} / \si{\pico\second}$ & - & 0.025 & - & - & 0.025 & - \\
\hline\hline
\end{tabular}
\end{table}

\section{AP potential for tungsten}
\label{sec::hybrid_tungsten_potential}

\subsection{EAM potential}
\label{sec::eam_tungsten_potential}
We used the EAM potential for W developed in Ref. \cite{eam_tungsten_2013} which is given by the embedding function
\begin{equation}
\xi(x) = a_1^\xi \sqrt{x} + a_2^\xi x^2\,,
\label{eq:eam_F_Marinica}
\end{equation}
the pair potential
\begin{equation}
\Phi(x) = \sum_{i=1}^{n^\Phi} a_i^\Phi \left(\delta_i^\Phi - x\right)^3 \Theta\left(\delta_i^\Phi -x\right)\,,
\label{eq::eam_Phi_Marinica}
\end{equation}
and the electron charge density
\begin{equation}
\zeta_0(x) = \sum_{i=1}^{n^\zeta} a_i^\zeta \left(\delta_i^\zeta - x\right)^3 \Theta\left(\delta_i^\zeta -x\right)\,,
\label{eq::eam_rho_Marinica}
\end{equation}
where $\Theta(x) = 1 \text{ for } x \geq 0, 0 \text{ for } x < 0$.
The electron charge density $\zeta_0(r)$ becomes negative for small $r$ and has a maximum at $r_\text{cut}^\zeta$.
Thus, a constant value of the electron charge density is used in form of
\begin{equation}
\zeta(x) = \begin{cases} \zeta_0(r_\text{cut}^\zeta) & \text{for } r \leq r_\text{cut}^\zeta, \\
                         \zeta_0(r)                 & \text{for } r >    r_\text{cut}^\zeta\end{cases}
\label{eq::tungsten_rho_up}
\end{equation}
to ensure a continuous derivative of the electron charge density.
Furthermore, the potential is extended to short range using the universal potential of Ref. \cite{Ziegler1985}.

\subsection{ACE potential}
\label{sec::ace_tungsten_potential}

The ACE parametrization for W was trained on a large dataset of DFT data obtained using the FHI-aims all-electron code~\cite{BLUM20092175}. The training structures included bulk as well as defective configurations. This parametrization has been extensively tested but remains to be a preliminary version. The final ACE model for W will be presented in a separate publication in near future.

\subsection{Combined EAM and ACE potential}
\label{sec::tungsten_hyb_potential}
The AP potential for W was constructed following the strategy described in Ref. \cite{adaptive_precision_potentials}.
\subsubsection{Optimizing the EAM potential}
\label{sec::optimising_eam_tungsten}
At first, we introduce an energy offset $\Delta \xi$ since the equilibrium energies of atoms differ between the EAM and the ACE description.
Thus, the embedding energy is given as
\begin{equation}
\xi^\text{Fit}(\zeta) = \xi(\zeta) + \Delta \xi\,.
\label{eq::tungsten_F_up}
\end{equation}
The EAM potential described by \cref{eq::eam_rho_Marinica,eq::eam_Phi_Marinica,eq::tungsten_F_up} is optimized using Atomicrex \cite{atomicrex}.
The minimized loss function is
\begin{equation}
\begin{split}
\mathcal{L}(\mathcal{A}) =& \sum_{s\in\mathcal{S}} \left\langle \left(\frac{E_{s,i}^\text{targ} - E_{s,i}^\text{pred}}{\delta^\text{tol}_{E_\text{md}}}\right)^2
+ \left(\frac{\lVert F_{s,i}^\text{targ} - F_{s,i}^\text{pred}\rVert}{\delta^\text{tol}_{F_\text{md}}}\right)^2 \right\rangle_i\\
& +\sum_{o\in\mathcal{O}} \left(\frac{A^\text{targ}_o - A^\text{pred}_o}{\delta^\text{tol}_o} \right)^2\,,
\label{eq::atomicrex_loss_function}
\end{split}
\end{equation}
where the notation $\langle x_i\rangle_i = \sum_{i=1}^{N}\frac{x_i}{N}$ is used.
The target values calculated with ACE and the used tolerances are given in \cref{tab::target_values_tungsten}.
The optimized parameters are shown in \cref{tab::parameters_eam_tungsten_potential}.

\begin{table}[tp]
\caption{\label{tab::target_values_tungsten}Target values with tolerance used in the loss function \cref{eq::atomicrex_loss_function} for the optimization of the EAM potential.}
\centering
\begin{tabular}{lrr}
\hline\hline
property                           & target value                                               & tolerance $\delta^\text{tol}$           \\
\hline
scalar properties $A_o$\\
lattice parameter $a_0^\text{BCC}$ & \round{4}{3.18395029539668}\,\si{\angstrom}                & \SI{0.001}{\angstrom}                   \\
cohesive energy $E_\text{coh}$     & \round{4}{-11.169362103864}\,\si{\electronvolt\per atom}   & \SI{0.1  }{\electronvolt\per atom}      \\
bulk modulus $B$                   & \round{4}{279.049352592162}\,\si{\giga\pascal}             & \SI{1    }{\giga\pascal}                \\
elastic constant $C_{11}$          & \round{4}{506.470248474139}\,\si{\giga\pascal}             & \SI{1    }{\giga\pascal}                \\
elastic constant $C_{12}$          & \round{4}{165.338904651174}\,\si{\giga\pascal}             & \SI{1    }{\giga\pascal}                \\
elastic constant $C_{44}$          & \round{4}{135.992410510025}\,\si{\giga\pascal}             & \SI{1    }{\giga\pascal}                \\
properties per structure $s$\\
force $F$                          & MD simulation                                              & \SI{0.01 }{\electronvolt\per\angstrom}  \\
potential energy $E$               & MD simulation                                              & \SI{0.01 }{\electronvolt}               \\
\hline\hline
\end{tabular}
\end{table}

\begin{table}[tb]
\caption[parameters of tungsten EAM potential]{Parameters of the optimized EAM potential at 300K.}
\label{tab::parameters_eam_tungsten_potential}
\begin{tabular}{lrlr}
\hline\hline
param & value                            & param & value                  \\
\hline
$r^\text{int}_\text{lo}$ & 1.10002200044 & $\delta^\Phi_8$ & 3.85904      \\
$r^\text{int}_\text{hi}$ & 2.10004200084 & $\delta^\Phi_9$ & 4.10323      \\
$Z$ & 74.0                               & $\delta^\Phi_{10}$ & 4.73354   \\
$\epsilon_0$ & 0.005526349406            & $\delta^\Phi_{11}$ & 4.8959    \\
$a_\text{ZBL}$ & 0.08702457248897934     & $\delta^\Phi_{12}$ & 5.09081   \\
$a^\Phi_1$ & 3663.54                     & $\delta^\Phi_{13}$ & 5.27739   \\
$a^\Phi_2$ & -3663.85                    & $\delta^\Phi_{14}$ & 5.40309   \\
$a^\Phi_3$ & 128.443                     & $\delta^\Phi_{15}$ & 5.45078   \\
$a^\Phi_4$ & 2.10845                     & $a^\rho_1$ & -267.394          \\
$a^\Phi_5$ & 5.10311                     & $a^\rho_2$ & 0.486763          \\
$a^\Phi_6$ & -4.06894                    & $a^\rho_3$ & -0.0425619        \\
$a^\Phi_7$ & 1.25081                     & $a^\rho_4$ & 0.0330214         \\
$a^\Phi_8$ & 1.64931                     & $\delta^\rho_1$ & 2.5          \\
$a^\Phi_9$ & -1.4248                     & $\delta^\rho_2$ & 3.1          \\
$a^\Phi_{10}$ & -0.761194                & $\delta^\rho_3$ & 3.5          \\
$a^\Phi_{11}$ & 1.93524                  & $\delta^\rho_4$ & 4.9          \\
$a^\Phi_{12}$ & -0.70151                 & $a^\xi_1$ & -6.24657      \\
$a^\Phi_{13}$ & 0.0935973                & $a^\xi_2$ & -0.0836583    \\
$a^\Phi_{14}$ & -1.40395                 & $\Delta \xi$ & -2.14333          \\
$a^\Phi_{15}$ & 1.13035                  & $b_0^\Phi$ & 15390.7852469191  \\
$\delta^\Phi_1$ & 2.74456                & $b_1^\Phi$ & -41436.4792651818 \\
$\delta^\Phi_2$ & 2.74451                & $b_2^\Phi$ & 45168.4307517119  \\
$\delta^\Phi_3$ & 2.28653                & $b_3^\Phi$ & -24693.6420961014 \\
$\delta^\Phi_4$ & 2.91147                & $b_4^\Phi$ & 6736.28696027297  \\
$\delta^\Phi_5$ & 2.96512                & $b_5^\Phi$ & -731.611626209140 \\
$\delta^\Phi_6$ & 3.07694                & $r^\rho_\text{cut}$ & 2.46396652011306 \\
$\delta^\Phi_7$ & 3.53116                & {}          & {}               \\
\hline\hline
\end{tabular}
\end{table}

\begin{figure}[tb]
\includegraphics[width=3.37in]{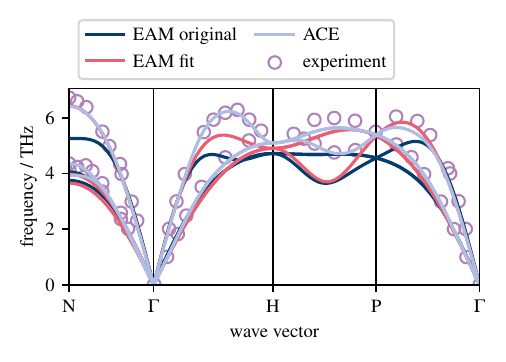}
\caption{\label{Fig::tungsten_fit_eam_phonon_spectrum}
Phonon spectra calculated with ASE\cite{atomic_simulation_enviroment} of the fitted EAM potential, original EAM potential, the ACE potential and experimental values\cite{CHEN196473} measured by inelastic neutron scattering at room temperature.
}
\end{figure}

Just like for the EAM original potential form Ref. \cite{eam_tungsten_2013}, the short-range interaction is given by the Coulomb energy screened by the ZBL screening function $\phi^\text{ZBL}$, namely
\begin{equation}
\Phi^\text{ZBL}(r) = \frac{1}{4\pi\epsilon_0} \frac{Z_1 Z_2 \si{\elementarycharge^2}}{r} \phi^\text{ZBL}(r)\,,
\end{equation}
with the vacuum permittivity $\epsilon_0\approx\SI{55.26e-4}{\elementarycharge^2\per\electronvolt\angstrom}$, the nuclear charge number $Z_{1,2}$ of the respective atom.
The ZBL screening function $\phi^\text{ZBL}$ is\cite{ZBL_potential}
\begin{equation}
\begin{split}
\phi^\text{ZBL} =& 0.1818e^{-3.2x^\text{ZBL}} + 0.5099e^{-0.9423x^\text{ZBL}}\\
&+ 0.2802e^{-0.4029x^\text{ZBL}} + 0.02817e^{-0.2016x^\text{ZBL}}
\end{split}
\end{equation}
with the reduced distance $x^\text{ZBL}=r/a^\text{ZBL}$ where
\begin{equation}
a^\text{ZBL} = \frac{0.8854 a_0}{Z_1^{0.23} + Z_2^{0.23}},
\end{equation}
where $a_0=\SI{0.529}{\angstrom}$ is the Bohr radius.
For W-W interaction with $Z_1=Z_2=74$ follows $a_\text{ZBL}\approx\SI{0.0870}{\angstrom}$.
The ZBL potential $\Phi^\text{ZBL}$ and the pair potential $\Phi$ are interpolated according to Ref. \cite{interpolation_eam_zbl} between $r^\text{int}_\text{lo}$ and $r^\text{int}_\text{hi}$ with the polynomial
\begin{equation}
\Phi^\text{int}(x) = \sum_{m=0}^{5} b_m^\Phi x^m
\end{equation}
so that the pair potential and its first two derivatives are continuous at $r^\text{int}_\text{lo}$ and $r^\text{int}_\text{hi}$.
Thereby we get the pair potential
\begin{equation}
\Phi^\text{Fit}(r) = \begin{cases} \Phi^\text{ZBL}(r) & \text{for } r < r^\text{int}_\text{lo}\,, \\
                        \Phi^\text{int}(r) & \text{for } r^\text{int}_\text{lo} \leq r \leq r^\text{int}_\text{hi}\,, \\
                        \Phi(r)   & \text{for } r^\text{int}_\text{hi} < r\,.\end{cases}
\label{eq::tungsten_Phi_up}
\end{equation}
Hence, we use the optimized EAM potential described by \cref{eq::tungsten_F_up,eq::tungsten_Phi_up,eq::tungsten_rho_up} with the parameters given in \cref{tab::parameters_eam_tungsten_potential}.
Basic properties calculated with the optimized EAM potential compared with the original EAM potential and the ACE potential are shown in \cref{tab::properties_tungsten_potentials}.
The elastic constants and related bulk properties correspond to the ACE reference values.
Properties not included in the Atomicrex optimization, such as surface energies or formation energies of interstitials, are not reproduced. The phonon spectra of the optimized EAM potential and the ACE potential compared in \cref{Fig::tungsten_fit_eam_phonon_spectrum} are in good agreement.

\begin{table}[tb]
\caption[properties of tungsten potentials]{Properties calculated with tungsten potentials.}
\label{tab::properties_tungsten_potentials}
\begin{tabular}{lrrr}
\hline\hline
{}                                                & {ACE}  & \multicolumn{2}{c}{EAM} \\
{}                                                & {}     & {fit} & {original}      \\
Interstitial Formation Energy / eV                                                   \\
100-dumbbell / eV                                 & 12.47  & 8.77   & 12.94          \\
tetrahedral / eV                                  & 11.03  & 8.66   & 10.43          \\
Elastic Constant C11 / GPa                        & 506    & 508    & 523            \\
Elastic Constant C12 / GPa                        & 165    & 162    & 203            \\
Elastic Constant C44 / GPa                        & 136    & 140    & 160            \\
Bulk Modulus / GPa                                & 279    & 277    & 310            \\
Shear Modulus 1 / GPa                             & 136    & 140    & 160            \\
Shear Modulus 2 / GPa                             & 171    & 173    & 160            \\
Poisson Ratio                                     & 0.25   & 0.24   & 0.28           \\
Lattice Constant / A                              & 3.1840 & 3.1837 & 3.1400         \\
Cohesive Energy / eV                              & -11.17 & -11.17 & -8.90          \\
Vacancy Formation Energy (bcc) / eV               & 3.36   & 4.01   & 3.49           \\
Surface Energy 111 / J/m$^2$                      & 3.62   & 3.55   & 2.96           \\
Surface Energy 100 / J/m$^2$                      & 4.11   & 3.10   & 2.72           \\
Surface Energy 110 / J/m$^2$                      & 3.51   & 2.95   & 2.31           \\
\hline\hline
\end{tabular}
\end{table}

\begin{table}[tb]
\caption{\label{tab::model_parameters_tungsten}Parameters of the adaptive-precision model and values of the parameters for tungsten at $\SI{300}{\kelvin}$.}
\begin{center}
\begin{tabular}{llllll}
\hline\hline
parameter & value & parameter & value & parameter & value \\
\hline
$N_\text{buffer}$          & 0\,atoms\phantom{\quad} & $\text{CSP}_\text{hi}$     & $\SI{1.6}{\angstrom^2}$             & $N_{\lambda,\text{avg}}$   & 110                          \\
$N_{\text{CSP},\text{avg}}$& 110                     & $r_{\lambda,\text{lo}}$    & \SI{4.0}{\angstrom}                 & $\Delta\lambda_\text{min}$ & 0.01 \\
$\text{CSP}_\text{lo}$     & $\SI{1.5}{\angstrom^2}$ & $r_{\lambda,\text{hi}}$    & \SI{12.0}{\angstrom}\phantom{\quad} & $|\Omega_i|$               & 800\,atoms                   \\
\hline\hline
\end{tabular}
\end{center}
\end{table}
\begin{figure*}[tbh]
\includegraphics[width=2.61in]{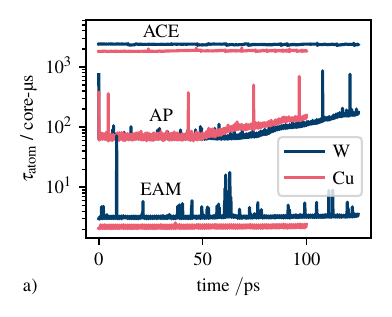}
\includegraphics[width=2.20in]{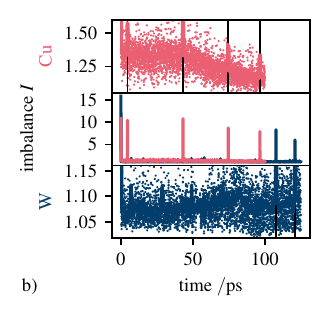}
\includegraphics[width=1.80in]{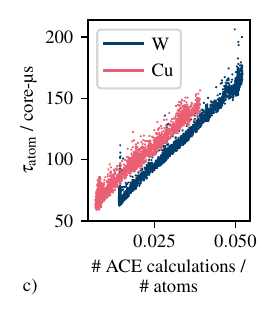}
\caption{\label{Fig::efficiency:over:time}
a) Compute time $\tau_\text{atom}$ per atom and time step according to \cref{eq::tau:atom} dependent on the simulated time for all nanoindentations.
b) Imbalance $I$ according to \cref{eq::imbalance} dependent on the simulated time for the AP simulations. The central plot shows that the system is initially and at restarts of the simulation unbalanced. The part of the imbalance-axis corresponding to a balanced system, i.e $I\approx1$, is enlarged at the top for Cu and at the bottom for W.
The time of restart is marked with vertical lines in the enlarged plots.
c) Correlation of $\tau_\text{atom}$ and the number of precisely calculated atoms. 
All measured values in c) and in the enlarged plots in b) are shown as points, while the a moving average of $\SI{1}{\pico\second}$-intervals is shown as line.
}
\end{figure*}
\subsubsection{Switching function}
\label{sec::switching_function_tungsten}
The parameters of the switching function $\lambda$ are set according to the strategy used in Ref. \cite{adaptive_precision_potentials}.
The used values are shown in \cref{tab::model_parameters_tungsten}.
The centro-symmetry parameter uses neighboring atoms which are on opposite positions of the central-atom and is usually calculated for the atoms of the nearest neighbor shell\cite{csp}.
The distance in a BCC lattice between the first and second neighbor shell is with $(1-\sqrt{3}/2)\approx0.13$ lattice constants relatively small, whereas the distance between the second and third neighbor shell is $(\sqrt{2}-1)\approx0.41$ lattice constants.
Therefore, one easily separate the third from the first two neighbor shells.
Thus, we calculate the CSP with 14 nearest neighboring atoms which corresponds to the number of neighbors in the first two neighbor shells.

\section{Computational efficiency}
\label{sec::computational:efficiency}

The compute time required for the simulation of the nanoindentations are reported in \cref{Fig::speedup_nanoindentation_copper,Fig::speedup_nanoindentation_tungsten} for the whole nanoindentation.
The automatic precision adjustment during the AP simulations affects also the compute time, which is why an analysis of the compute time dependent on the progress of the simulation provides further insights.
LAMMPS measures the number of simulated timesteps per wall-time second.
The compute time $\tau_\text{atom}$ per atom and timestep is given as
\begin{equation}
    \tau_\text{atom} = \frac{\text{\# processors}}{\text{\# timesteps per wall-time second} \cdot \text{\# particles}}\,,
    \label{eq::tau:atom}
\end{equation}
which is shown in \cref{Fig::efficiency:over:time}(a) dependent on the simulated time.
$\tau_\text{atom}$ is approximately constant for EAM and ACE for the whole simulation, while $\tau_\text{atom}$ starts to increase during the AP simulations.
The number of precisely calculated atoms increases during the simulation, as the dislocation density $\rho^\text{dxa}$ (\cref{eq::dislocation_density}) increases (cf. \cref{Fig::dislocation_density_tungsten,Fig::dislocation_density_copper}).
As the number of precisely calculated atoms and $\tau_\text{atom}$ are positively correlated (cf. \cref{Fig::efficiency:over:time}(c)), the increase of $\tau_\text{atom}$ over time is expected.

The imbalance $I$ is defined in LAMMPS as \cite{lammps_fix_balance}
\begin{equation}
    I = \frac{\text{max}\left\{\tau_p^\text{force} \right\}}{\langle\tau_p^\text{force}\rangle_p}\,,
    \label{eq::imbalance}
\end{equation}
where $\tau_p^\text{force}$ denotes the on processor $p$ measured force-calculation time.
$I=1$ applies for a perfectly balanced system, while $I>1$ indicates load imbalances.
The mean imbalance in the adaptive-precision simulations of W and Cu is $1.09$ and $1.29$, whereas the visualization dependent on the simulated time in \cref{Fig::efficiency:over:time}(b) shows for Cu a decrease of the imbalance when $\tau_\text{atom}$ increases.
Hence, the increase of $\tau_\text{atom}$ in the AP simulations is independent of the imbalance for W and weakened by the imbalance for Cu.
Furthermore, the fixing of the switching parameter in the zones of interest, as described in \cref{sec::adaptive_precision_potential}, results in a decrease of the imbalance of 0.12 in the AP Cu simulation compared to the AP Cu simulation in Ref. \cite{adaptive_precision_potentials}.
Thermal fluctuations of atoms affect the centro-symmetry parameter, which is used to calculate the switching parameter.
As discussed in Ref. \cite{adaptive_precision_potentials}, an atom, which is only due to thermal fluctuations detected for a precise calculation, causes due to the spatial transition zone the computation of ACE also for neighboring atoms and, thereby, increases the imbalance.
The existence of such atoms at room temperature is less likely for W than for Cu due to the higher melting point of W.
Therefore, the imbalance $I$ is smaller for W than for Cu in the AP nanoindentations (cf. \cref{Fig::efficiency:over:time}(b)).
As the number of fast calculated atoms decreases over time, in particular through the gliding of prismatic dislocation loops (cf. \cref{Fig::dislocation_networks_copper}), the number of atoms, that are susceptible to incorrect precise treatment, and, thus, the imbalance decrease at the end of the AP Cu simulation.

\section{Tungsten nanoindentation with optimized EAM potential}
\label{sec::tungsten_nanoindentation_optimised_eam}

\begin{figure}[tb]
\begin{tikzpicture}[x=1.000in,y=0.86in]
  \node[anchor=north west,inner sep=0] at (0,0) {
    \includegraphics[width=1in]{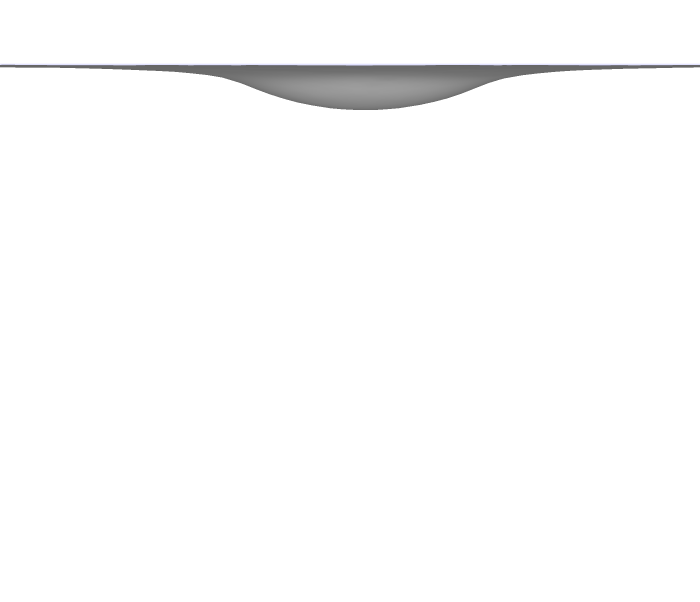}
  };
  \node[anchor=south] at (0.5,0.0) {$h/a_\text{c}\approx0.41$};
  \node[anchor=north west] at (0,-0.5) {\textcolor{black}{a)}};
  \node [coordinate] (tripod_origin) at (0.85,-0.80) [] {};
  \draw[->] (tripod_origin) -- ($(tripod_origin) - (0.1,0)$) node[anchor=east]  {$[100]$};
  \draw[->] (tripod_origin) -- ($(tripod_origin) + (0,0.166)$) node[anchor=south] {$[001]$};

  \node[anchor=north west,inner sep=0] at (1.1,0) {
    \includegraphics[width=1in]{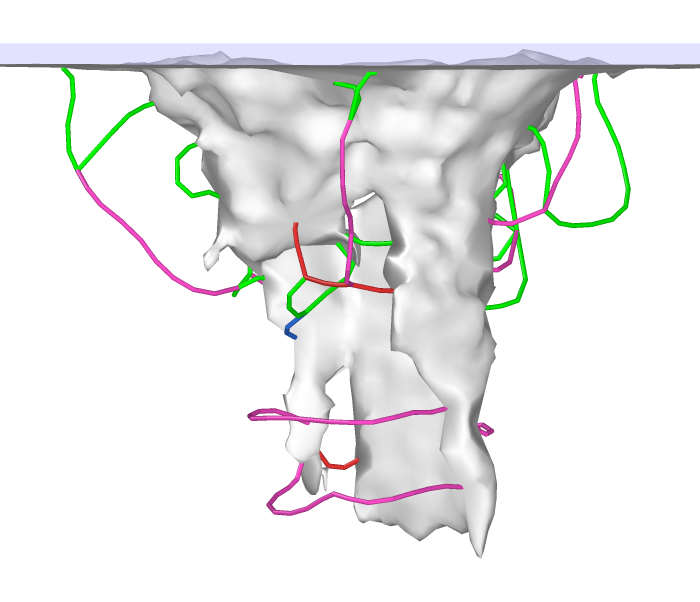}
  };
  \node[anchor=south] at (1.6,0.0) {$h/a_\text{c}\approx0.62$};
  \node[anchor=north west] at (1.1,-0.5) {\textcolor{black}{b)}};
  \draw[|-|] (2.15,0) -- (2.15,-1) node[midway, above, rotate=270] {$\SI{115.0}{\angstrom}$};

  \node[anchor=north west,inner sep=0] at (0,-1) {
    \includegraphics[width=1in]{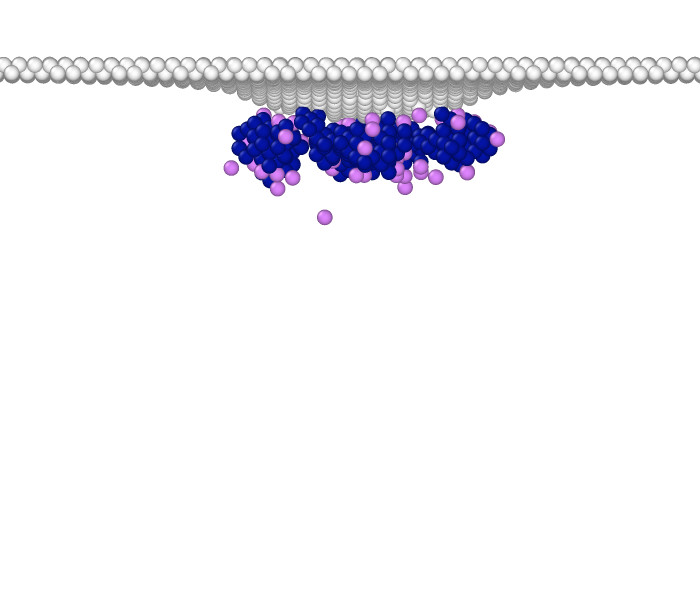}
  };
  \node[anchor=north west] at (0,-1.5) {\textcolor{black}{c)}};
  \node [coordinate] (tripod_origin) at (0.85,-1.80) [] {};
  \draw[->] (tripod_origin) -- ($(tripod_origin) - (0.1,0)$) node[anchor=east]  {$[100]$};
  \draw[->] (tripod_origin) -- ($(tripod_origin) + (0,0.166)$) node[anchor=south] {$[001]$};

  \node[anchor=north west,inner sep=0] at (1.1,-1) {
    \includegraphics[width=1in]{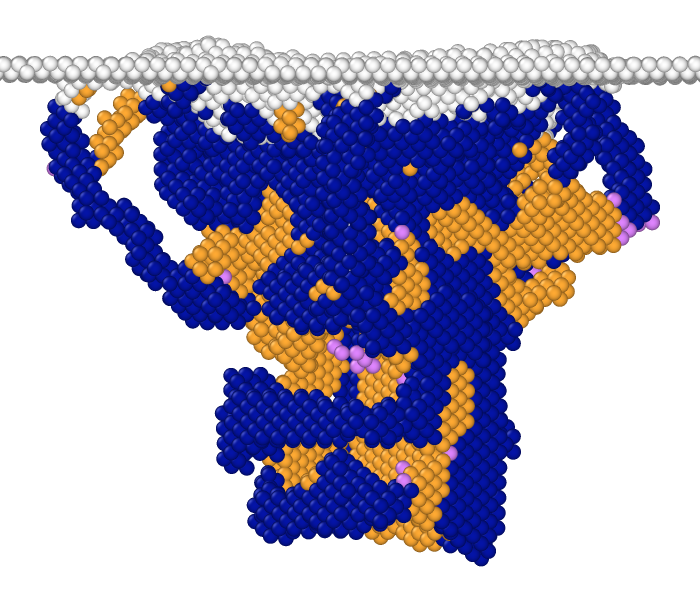}
  };
  \node[anchor=north west] at (1.1,-1.5) {\textcolor{black}{d)}};
  \draw[|-|] (2.15,-1) -- (2.15,-2) node[midway, above, rotate=270] {$\SI{115.0}{\angstrom}$};

  \node[anchor=north west,inner sep=0] at (0,-2) {
    \includegraphics[width=1in]{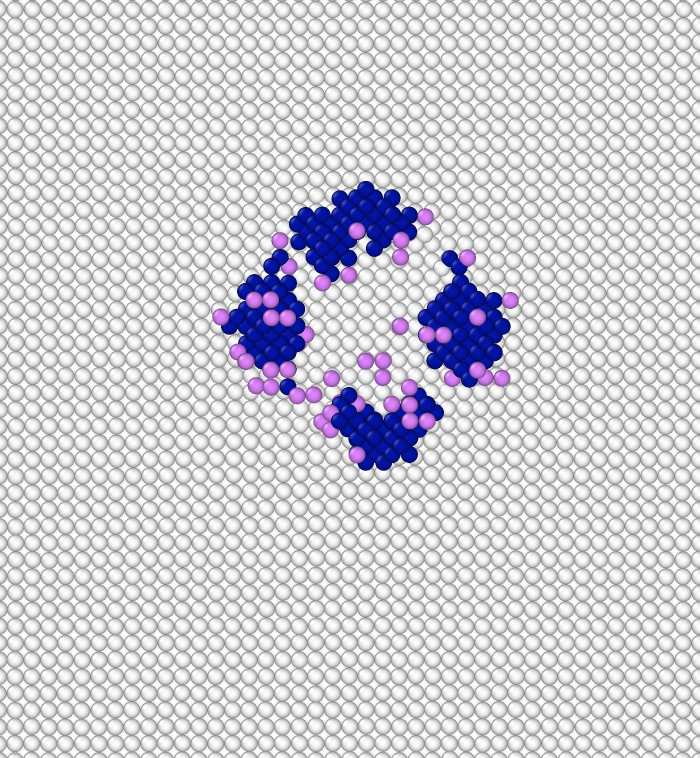}
  };
  \node[anchor=north west] at (0,-2) {\textcolor{black}{e)}};
  \node [coordinate] (tripod_origin) at (0.85,-2.15) [] {};
  \draw[->] (tripod_origin) -- ($(tripod_origin) - (0.1,0)$) node[anchor=east]  {$[100]$};
  \draw[->] (tripod_origin) -- ($(tripod_origin) - (0,0.166)$) node[anchor=north] {$[0\overline{1}0]$};

  \node[anchor=north west,inner sep=0] at (1.1,-2) {
    \includegraphics[width=1in]{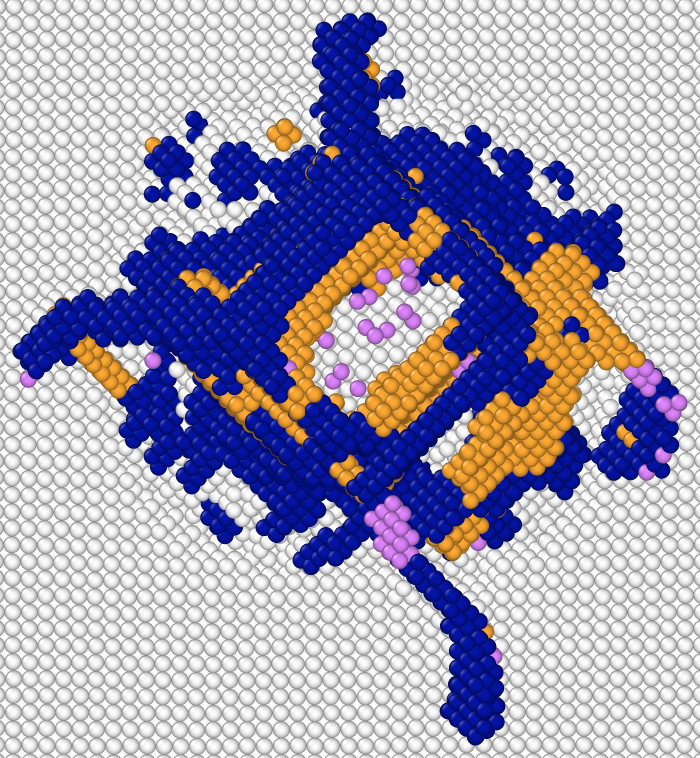}
  };
  \node[anchor=north west] at (1.1,-2) {\textcolor{black}{f)}};
  \draw[|-|] (2.15,-2) -- (2.15,-3.259) node[midway, above, rotate=270] {$\SI{144.6}{\angstrom}$};

  \node[anchor=north west,inner sep=0] at (0.0,-3.259) {
    \includegraphics[width=1in]{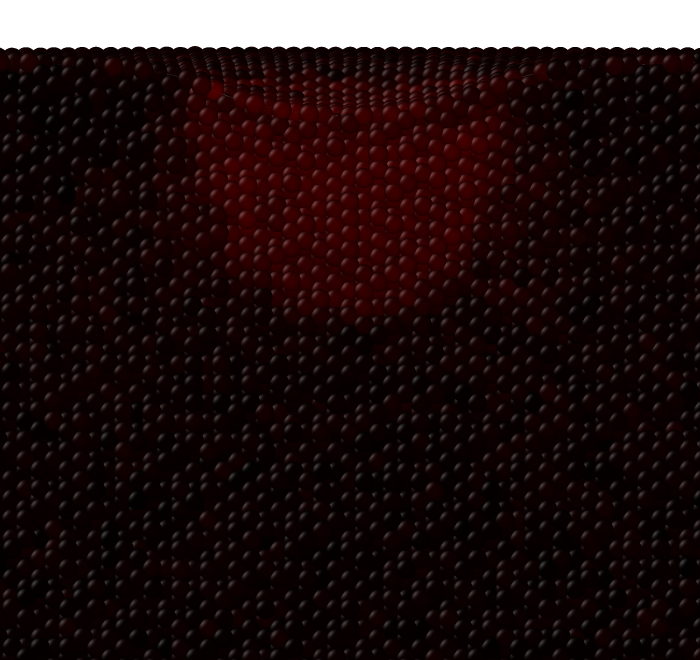}
  };
  \node[anchor=north west] at (0,-4.1) {\textcolor{white}{g)}};
  \draw[->,white] (0.4,-4.2) -- (0.9,-4.2) node[midway, above] {$[110]$};

  \node[anchor=north west,inner sep=0] at (1.1,-3.259) {
    \frame{\includegraphics[width=1in]{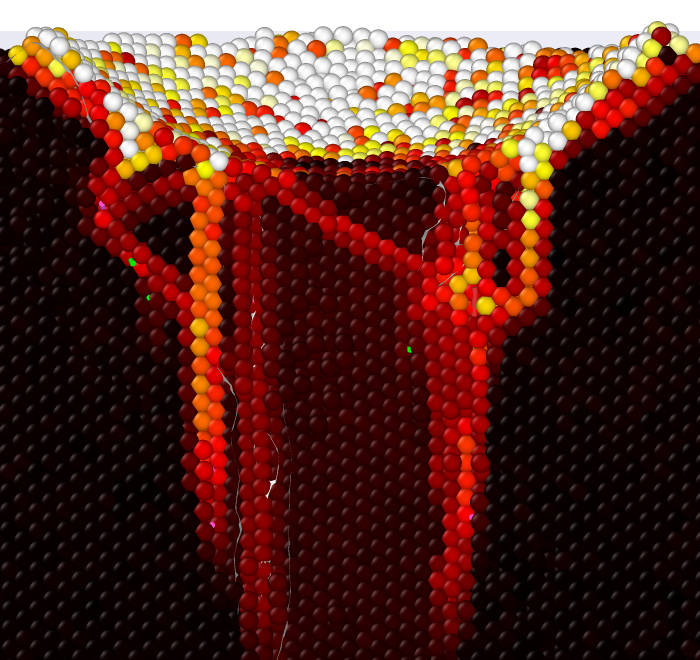}}
  };
  \node[anchor=north west] at (1.1,-4.1) {\textcolor{white}{h)}};
  \draw[|-|] (2.15,-3.259) -- (2.15,-4.355) node[midway, above, rotate=270] {$\SI{105.8}{\angstrom}$};

  \node[anchor=north west,inner sep=0, rotate=270] at (2.6,-3.259) {
    \frame{\includegraphics[width=0.94in]{Tungsten_plots_prod_shear_strain_color_bar.png}}
  };
  \draw[|-|] (2.6,-3.259) -- (2.6,-4.355) node[midway, above, rotate=270] {shear strain};
  \node[anchor=west] at (2.6,-3.259) {0};
  \node[anchor=west] at (2.6,-4.335) {1.5};

  \draw[]                         (2.4,-1.2)                 node[anchor=west] {\underline{BDA}};
  \draw[black,fill=bda_white]     (2.4,-1.4) circle (0.10cm) node[anchor=west] {\,surface};
  \draw[black,fill=bda_lightblue] (2.4,-1.6) circle (0.10cm) node[anchor=west] {\,vacancy};
  \draw[black,fill=bda_darkblue]  (2.4,-1.8) circle (0.10cm) node[anchor=west] {\,non-screw disl.};
  \draw[black,fill=bda_orange]    (2.4,-2.0) circle (0.10cm) node[anchor=west] {\ twin/screw disl.};
  \draw[black,fill=bda_green]     (2.4,-2.2) circle (0.10cm) node[anchor=west] {\ \{110\}planar fault};
  \draw[black,fill=bda_purple]    (2.4,-2.4) circle (0.10cm) node[anchor=west] {\ unidentified};

  \node[anchor=west] at (2.4, -0.0) {\underline{DXA}};
  \node[anchor=west] at (2.4, -0.2) {\textcolor{dxa_bcc_green}{\rule{0.2cm}{0.2cm}} $1/2\langle 111\rangle$};
  \node[anchor=west] at (2.4, -0.4) {\textcolor{dxa_bcc_pink}{ \rule{0.2cm}{0.2cm}} $\langle 100\rangle$};
  \node[anchor=west] at (2.4, -0.6) {\textcolor{dxa_bcc_blue}{ \rule{0.2cm}{0.2cm}} $\langle 110\rangle$};
  \node[anchor=west] at (2.4, -0.8) {\textcolor{dxa_bcc_red}{  \rule{0.2cm}{0.2cm}} other};
\end{tikzpicture}
\caption{\label{Fig::bda_networks_tungsten_eam}
Tungsten nanoindentation simulated with the optimized EAM potential (\cref{tab::parameters_eam_tungsten_potential}) visualized with a-b) the dislocation extraction algorithm (DXA) and c-f) the BCC defect analysis (BDA).
g-h) A cut in the $(1\overline{1}0)$ plane through the contact point is used to visualize the by OVITO calculated shear strain.
Scale bars are given on the right side of each row.
}
\end{figure}

To validate whether the nanoindentation simulations with the optimized EAM potential would give results similar to the reference ACE potential, we performed additional simulation runs. Dislocation lines, defects and the shear strain from these simulations are visualized in \cref{Fig::bda_networks_tungsten_eam} for the same indentation depths as in \cref{Fig::dislocation_networks_tungsten,Fig::bda_networks_tungsten}. 

For the normalized indentation depth $h/a_\text{c}\approx 0.41$ (left column), there are primarily non-screw dislocations detected by the BDA algorithm for the original EAM potential (cmp. \cref{Fig::bda_networks_tungsten}(g) as well as for the optimized EAM potential (cmp. \cref{Fig::bda_networks_tungsten_eam}(e). Unlike in the AP/ACE simulations, no twins are detected.

At the maximum indentation depth (right column), we observe less pile-up when the optimized EAM potential is used than when the original EAM potential is used (cmp. \cref{Fig::bda_networks_tungsten_eam}(d) and \cref{Fig::dislocation_networks_tungsten}(a).
Furthermore, twins are detected by BDA in the optimized-EAM nanoindentation (cmp \cref{Fig::bda_networks_tungsten_eam}(d) while there are none in the original-EAM nanoindentation.
However, there are still numerous $\langle 100\rangle$ dislocation lines below the indenter (cmp. \cref{Fig::bda_networks_tungsten_eam}(b) and the shear strain in higher only below the indenter for both EAM potentials (cmp. \cref{Fig::bda_networks_tungsten_eam}h and \cref{Fig::dislocation_networks_tungsten}j), in contrast to the observed $1/2\langle 111\rangle$ loops observed in the AP/ACE nanoindentation.

In conclusion, the optimized EAM potential predicts a less pronounced pile-up at the indent's rim and the existence of twins, unlike the original EAM potential, which somewhat resemble the AP/ACE results. However, the dislocation mechanisms remain the same as for the original EAM potential. Therefore, the mechanisms observed in the AP/ACE nanoindentation simulations cannot be simply achieved via optimization of the EAM potential.

\bibliography{refs.bib}

\end{document}